\documentclass[traditabstract]{aa}
\usepackage{graphicx}
\usepackage{epsfig}
\usepackage{color}
\usepackage{txfonts}
\usepackage{natbib}
\usepackage{multirow}
\bibpunct{(}{)}{;}{a}{}{,} 

\begin{document}

\title{Minor mergers and their impact on the kinematics of
  old and young stellar populations in disk galaxies}
\author{Yan Qu\inst{1}, Paola Di Matteo\inst{1}, Matthew D. Lehnert\inst{1}, Wim van Driel\inst{1}, \and Chanda J. Jog\inst{2}}

\authorrunning{Qu et al.}

\institute{GEPI, Observatoire de Paris, CNRS, Universit\'e
  Paris Diderot, 5 place Jules Janssen, 92190 Meudon, France\\
\email{yan.qu@obspm.fr}
\and
  Department of Physics, Indian Institute of Science, Bangalore
  560012, India}

\date{Accepted, Received}

\abstract{By means of N-body simulations we investigate the impact
of minor mergers on the angular momentum and dynamical properties of
the merger remnant.  Our simulations cover a range of initial
orbital characteristics and gas-to-stellar mass fractions (from 0 to
$20\%$), and include star formation and supernova feedback. We
confirm and extend previous results by showing that the specific
angular momentum of the stellar component always decreases
independently of the orbital parameters or morphology of the
satellite, and that the decrease in the rotation velocity of the
primary galaxy is accompanied by a change in the anisotropy of the
orbits. However, the decrease affects only the old stellar
population, and not the new population formed from gas during the
merging process. This means that the merging process induces an
increasing difference in the rotational support of the old and young
stellar components, with the old one lagging with respect to the
new. Even if our models are not intended specifically to reproduce
the Milky Way and its accretion history, we find that, under certain
conditions, the modeled rotational lag found is compatible with that
observed in the Milky Way disk, thus indicating that minor mergers
can be a viable way to produce it. The lag can increase with the
vertical distance from the disk midplane, but only if the satellite
is accreted along a direct orbit, and in all cases the main
contribution to the lag comes from stars originally in the primary
disk rather than from stars in the satellite galaxy. We also discuss
the possibility of creating counter-rotating stars in the remnant
disk, their fraction as a function of the vertical distance from the
galaxy midplane, and the cumulative effect of multiple mergers on
their creation.}


\maketitle

\section{Introduction}

Rotationally supported disks account for only a small fraction of the
mass in the local Universe, but they contain most of the angular momentum
(hereafter AM). The way disks acquire and redistribute their AM represents
one of the most challenging problems for models of galaxy formation and
evolution. According to the current cosmological paradigm, baryons and
dark halos in galaxies acquire their spin through tidal torques exerted
by adjacent structures at early times. This AM is then redistributed
among the different galaxy components through a number of internal and
external processes as the galaxy evolves. Among the internal processes,
bars, lopsidedness, spiral patterns and other coherent structures are
efficient in redistributing AM in galaxies, as many studies have shown
\citep[e.g.,][]{athanassoula205,debattista206,minchev210}. These stellar
asymmetries can be stimulated or strengthened by external processes, such
as accretion of a few M$_{\odot}$ yr$^{-1}$ of gas from cosmological filaments
\citep[see][]{bournaud205a,bournaud205b} or tidal interactions and
mergers \citep{jogM206,mapelliMBH208,reichard209}.  In particular,
during an interaction orbital AM is converted into internal rotation,
in an outside-in manner: the components which first interact are the
most extended ones, while the more tightly bound components experience
strong tidal effects only in the final phases of the merging process
\citep{barnes192,dimatteo208a,dimatteo209}.

Many studies have shown that major mergers have a catastrophic
impact on the ordered motion of the pre-existing galaxies. If the
progenitors have disks, they are usually destroyed by the strong
energy and AM redistribution taking place during the interaction
\citep{toomre177,bendo200, naabB203,bournaudJC205,jesseit209}, unless
peculiar orbital configurations are chosen \citep{puerari201,crocker209}.
The fraction of gas present in the progenitor disks can also
influence the morphology and kinematics of the final remnant
\citep[e.g.,][]{hopkins209}, but this depends as well on the gas physics
implemented in the models \citep[e.g.,][]{bournaud210}.  In the case of
pressure-supported progenitors, in turn, the tidal torques exerted by
the companion can be strong enough to produce high rotational support
($v/\sigma>1$) at large radii, even in merger remnants having an
elliptical-like morphology \citep{dimatteo209}. More attention has
been given to the study of the impact of AM redistribution in major
mergers than in minor mergers (with mass ratios $\le$0.1). This
despite the fact that minor mergers are expected to be much more
common than major mergers \citep{fakhouriM208} and that many traces
of ongoing or past interactions are visible both in the Milky Way
\citep[see][for a recent review]{klement210}, our neighbor galaxy
Andromeda \citep{ibata201,mcconnachie209} and other galaxies in the local
Universe \citep{martinez-delgado210}. \citet{dimatteo210} pointed out
that the AM redistribution during minor mergers also have an impact on
the kinematics of stellar disks and may explain the distribution of the
orbital eccentricities of stars in the solar neighborhood. Understanding
how AM is redistributed during such episodes is fundamental to the
understanding of how disks can be maintained, how their kinematics can be
affected, and what the signatures are of these processes on the dynamical
properties of different stellar populations.

This paper is the second of a series where we study, by means of
numerical simulations, the impact of minor mergers on AM
redistribution in galaxies. In \cite{quDM210a} (hereafter Paper I),
we investigated the impact of dissipationless minor mergers on disk
galaxies, showing in particular that the initially non-rotating dark
matter halo of the primary galaxy always gains AM and that the
specific AM of the stellar component always decreases. We also
showed that this decrease in AM is accompanied by a change in
stellar velocity anisotropy as the stellar orbits become less
tangentially dominated as the merger advances.  In this paper, we
aim to advance this analysis by studying simulations of dissipative
minor mergers, exploring a range of gas fractions and morphological
parameters for the primary galaxy and the satellite.  Star formation
and feedback from supernovae explosions are included in the models,
and we are able to trace the AM redistribution of all the galaxy
components: dark matter, gas, old stars (i.e., those already in the
galaxies before the interaction starts) and new stars (i.e., those
formed from the gas during the interaction). In particular, we aim
to understand if dissipative minor mergers still slow down the
stellar disk of the primary galaxy and if stellar populations of
different ages show a different AM content and different dynamical
properties in the final (i.e., post-merger) disk.

The paper is organized as follows: the numerical code, the initial
galaxy models and orbital conditions adopted for the runs are described
in \S~\ref{model}. Section~\ref{results} presents the main results,
in particular how the AM content of gas and the old and new stellar
populations is affected by a single and by two consecutive (\S~\ref{angm})
minor mergers, and how the rotational lag induced by this redistribution
affects the old stellar component (\S~\ref{rotlag}). The contribution
of accreted stars to the rotational lag, and the possibility to
distinguish them on the basis of their AM content, are discussed in
\S~\ref{discriminate}. Section~\ref{counter} discusses the effect
of single and two consecutive retrograde mergers on the fraction
of counter-rotating stars.  Finally, \S~\ref{discussion} presents a
discussion of the results found and the main conclusions.

\section{Models}\label{model}

All 121 simulations described in this paper are part of the GalMer
project\footnote{\emph{http://galmer.obspm.fr}} and were fully
described in \citet{chilingarianDMC209}. Here we recall the main
characteristics of the adopted galaxy models and orbital parameters,
as well as those of the numerical code employed to run the
simulations.

\begin{table}
\caption[]{Parameters for the initial models of the bulge, halo and
stellar and gaseous disks.}
\label{morphtable} \centering
\begin{tabular}{lccccccc}
\hline\hline & gS0 &  gSa & gSb & dE0 & dS0 & dSa & dSb \\ \hline
M$_{\rm B}\ [2.3\times 10^9 M_{\odot}]$ & 10. & 10. & 5. & 7. & 1 & 1 & 0.5\\
M$_{\rm H}\ [2.3\times 10^9 M_{\odot}]$ & 50.  & 50. & 75. & 3. & 5 & 5 & 7.5 \\
M$_{*}\ [2.3\times 10^9 M_{\odot}]$ & 40.  & 40. & 20. & 0 & 4 & 4 & 2 \\
M$_{\rm gas}/M_{*}$ & -- & 0.1 & 0.2 & -- & -- & 0.1 & 0.2\\
r$_{\rm B}\ [\mathrm{kpc}]$ & 2. &  2. &1. & 1.3 & 0.6 & 0.6 & 0.3\\
r$_{\rm H}\ [\mathrm{kpc}]$ & 10. & 10. & 12.& 2.2 & 3.2 & 3.2 & 3.8\\
$a_{*}\ [\mathrm{kpc}]$ & 4. &  4.& 5.& -- & 1.3 & 1.3 & 1.6\\
$h_{*}\ [\mathrm{kpc}]$ & 0.5 &  0.5 & 0.5& -- & 0.16 & 0.16 & 0.16\\
$a_{\rm gas}\ [\mathrm{kpc}]$ & -- &  5.& 6.& -- & -- & 1.6 & 1.9\\
$h_{\rm gas}\ [\mathrm{kpc}]$ & -- &  0.2 & 0.2 & -- & -- & 0.06 & 0.06\\ \hline
\end{tabular}
\end{table}

In this paper, we study the interaction and successive merger of a
satellite galaxy with a ten times more massive primary disk galaxy. Both
the primary and the satellite consist of a spherical non-rotating dark
halo and a central bulge, both modeled by Plummer spheres of masses
M$_{\rm H}$ and M$_{\rm B}$, respectively, and core radii r$_{\rm H}$
and r$_{\rm B}$, a stellar and an optional gas disk, represented by
Myamoto-Nagai density profiles of masses M$_{*}$ and M$_{\rm gas}$,
disk scale lengths a$_{*}$ and a$_{\rm gas}$, and scale heights h$_{*}$
and h$_{\rm gas}$, respectively. Both the primary and the satellite
galaxies span a range of morphologies, with a variety of bulge to
disk ratios, and gas-to-stellar mass fractions (from f$_{\rm gas}=0$
to f$_{\rm gas}=20\%$). Moving from early to late type systems,
we refer to these models respectively with the nomenclature gS0, gSa and
gSb for the primary galaxy, and dS0, dSa and dSb for the satellite. For
the satellites, we also consider a simple spheroid dominated model,
consisting only of a dark halo and a stellar bulge, without any disk
component -- we refer to this model as dE0.  All the parameters of the
galaxies described above are given in Table~\ref{morphtable}, and the
number of particles adopted to describe the different galaxy components
in Table~\ref{numbers}.

As we have done in \citet{quDM210a}, all galaxy models were first
evolved in isolation for 1~Gyr before the interaction starts, to let
the initial system reach a stable configuration. Once relaxed, the
two galaxies were placed at an initial distance of 100~kpc, with a
variety of relative velocities, to simulate different orbits (see
\citealp{chilingarianDMC209}, Table 9, for the orbital initial
conditions). To study the effect of multiple mergers, we also ran
some simulations in which the primary galaxy accretes consecutively
two or three satellites over a period of 3-5~Gyr. We also compared 
the dynamical properties of disks after minor mergers to those
heated by internal processes, such as scattering by massive clumps
formed in an initially unstable disk \citep{bournaud209}. For this,
we analyzed a simulation of a gas-rich, unstable disk galaxy from
\citet{dimatteo208b}. It uses a total number of 120,000 particles,
equally distributed among gas, stars and dark matter, and has the
same parameters as the gSb model (see Table~\ref{morphtable}) except
for the gas mass fraction, which is initially much higher, 50$\%$.

All the simulations were run using the Tree-SPH code described in
\citet{semelinC202}.  Gas is treated as isothermal, at a fixed
temperature of T=10$^4$~K. Prescriptions for star formation and
feedback from supernovae explosions are also included. The rate of
star formation is governed locally by a volume density Schmidt law
$\dot{{\rm M}_{\rm gas}}\propto \rho^{\alpha}{\rm M}_{\rm gas}$ with
$\alpha$=1.5, which is implemented in the code by means of a hybrid
particle approach (see \citealp{mihosH194, chilingarianDMC209} for
details). At each time step, a hybrid particle is characterized by
two mass values: the first is its total mass $M_{i}$ which stays
constant during the simulation and is used to calculate the
gravitational force, and the second is given by the gas content of
the particle, M$_{\rm i,gas}$, which changes with time according to
the Schmidt law and is used to evaluate hydrodynamical quantities.
At each time step the mass of the new stars, i.e. those formed since
the beginning of the simulation, is given by M$_{\rm i}-$M$_{\rm
gas}$.  If M$_{\rm i,gas}$ drops below 5$\%$ of the initial value,
the ``hybrid'' particle is considered to be totally converted into a
star-like particle and its small residual amount of gas spread out
over its (hybrid) neighbors. Note that in the hybrid particle
scheme, new stars follow the gas kinematics until the moment when
the hybrid particle they reside in is completely converted into
stars. This can represent a limitation of our approach, in the sense
that it does not allow us to properly follow the heating with time
of the newly formed stars by secular processes. Finally, for the
evaluation of the gravitational forces a softening length
$\epsilon$=200~pc is employed. The equations of motion are
integrated using a leap-frog algorithm, with a fixed time step of
0.5~Myr.  With these choices, the relative error in the conservation
of the total energy is close to 10$^{-6}$ per time step.

The nomenclature adopted for our simulations consists of a string of 13
characters: the first three (gS0, gSa or gSb) for the primary galaxy
and the following three for the satellite galaxy (dE0, dS0, dSa or
dSb), followed by the encounter identification string \citep[see][Table
9]{chilingarianDMC209} and the orientation of the disk of the primary
galaxy with respect to the orbital plane (33 or 60 degrees). Multiple
mergers are indicated by two more characters -- tn, where n is the number
of satellites accreted by the primary galaxy. For example, the string
gSadSat202dir33 refers to a gSa galaxy accreting two dwarf dSa galaxies,
whose initial orbital parameters are those corresponding to the string
02dir in Table 9 of \citet{chilingarianDMC209}.

\begin{table}
\caption[]{Particle numbers for primary disk galaxies and satellites.}
\label{numbers}
\begin{tabular}{lccc}
\hline\hline
 & gS0  & gSa & gSb \\ \hline
N$_{\rm gas}$  & --   &  80,000 & 160,000 \\
N$_{\rm star}$ & 320,000 & 240,000 & 160,000 \\
N$_{\rm DM}$   & 160,000 & 160,000 & 160,000 \\ \hline
\end{tabular}\\

\vspace{0.5cm}
\begin{tabular}{lcccc}
\hline\hline
 & dE0 & dS0 & dSa & dSb \\ \hline
N$_{\rm gas}$  & --  & -- & 8,000  & 16,000\\
N$_{\rm star}$ & 32,000 & 32,000 & 24,000 & 16,000\\
N$_{\rm DM}$   & 16,000 & 16,000 & 16,000 & 16,000\\ \hline
\end{tabular}
\end{table}

\section{Results}\label{results}

\begin{figure*}
\centering
\includegraphics[width=15cm,angle=0]{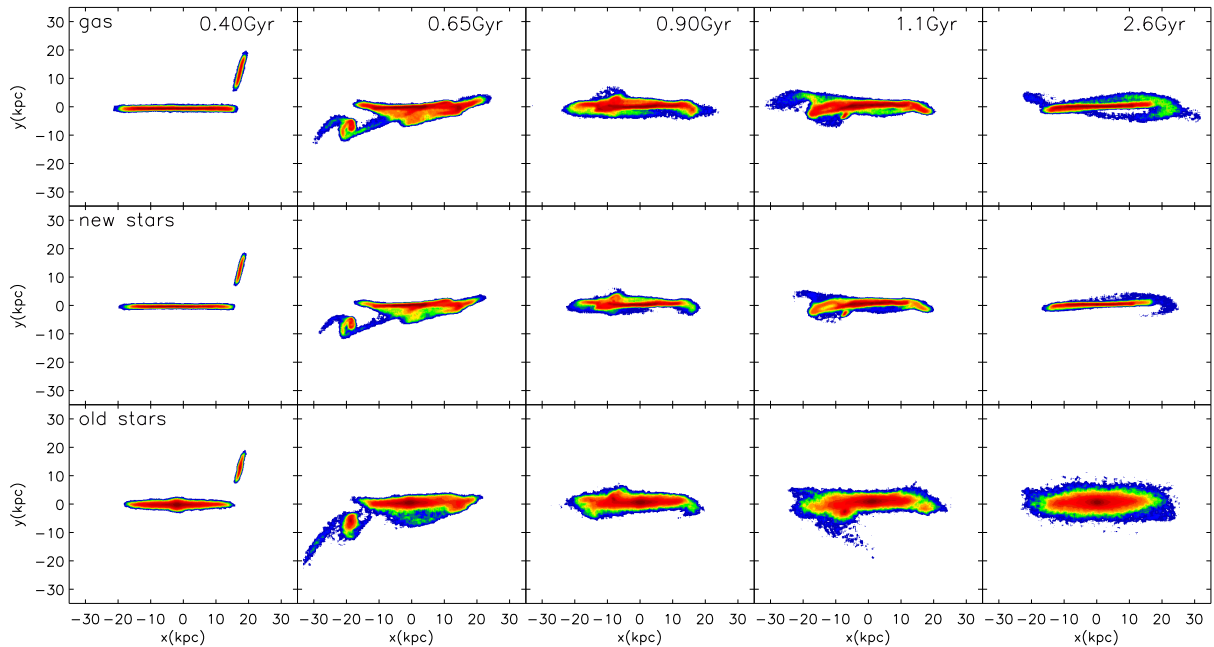}
\vspace{0.7cm}
\caption{\emph{From left to right}: evolution with time of gas (\emph{upper panels})
and stars (\emph{middle and lower panels}) during a minor merger between
a gSb and a dSb galaxy on a direct orbit (gSbdSb01dir33). The stellar component
is separated into new stars (formed during the interaction,
\emph{middle row of panels}), and old stars (those already present in
the galaxies before the interaction, \emph{lower panels}). In this projection the
disk of the primary galaxy is viewed edge-on.}
\label{all-3}
\end{figure*}

In Paper I, we showed that during dissipationless minor mergers
the orbital AM is redistributed into internal AM of the interacting
galaxies in an outside-in manner: the initially non-rotating dark halos
of primary and satellite start to acquire AM first, just after the first
pericenter passage between the two galaxies, while the strongest changes
in the primary stellar disk take place mostly in the final phase of the
collision, when the satellite is close to merging. All our simulations
also showed that the stellar disk loses part of its initial AM.  The aim
of this section is to investigate if these results are also valid for
dissipative mergers, if differences can be found in the rotational support
of the different stellar populations in the primary disk once the merger
is completed and what their main kinematic characteristics are.

\subsection{Gas, old stars, new stars and the evolution of their angular
momenta}\label{angm}

As discussed in \S~\ref{model}, our model galaxies contain both
``old stars'', which were already present before the start of the
interaction, and ``new stars'' formed from gas during the
interaction.  We will follow the nomenclature of old and new stars
throughout the text. An example of the spatial distribution and
morphological evolution of these two stellar components, as well as
the dissipative gas component, is shown in Fig.\ref{all-3}. Before
the first close passage of the satellite, which occurs at 0.5~Gyr,
gas and new stars are distributed in a morphologically thin and
dynamically cold disk, and old stars in a thicker and hotter
component and in a central bulge. At 0.65~Gyr, the primary disk
shows signs of a tidal perturbation: some gas and new stars are
clearly displaced from the midplane of the primary, and the old
stellar disk starts to thicken even further. Note also that the
satellite morphology has already been strongly modified at this
time. Signs of tidal perturbation of the primary galaxy are visible
all along the merging sequence after the first close passage of the
satellite. At 2.6~Gyr, or more than 1~Gyr after the merger ended,
its morphology has clearly been profoundly modified: a thick disk
made of old stars has been formed \citep[see ][for a description of
its morphological properties]{quDM211b} whereas gas and new stars
are still distributed in a thin disk component with some
perturbations at large radii, such as a warp and a tidal tail.

\begin{figure}
\vspace{0cm}\hspace{-0.2cm}
\includegraphics[width=9.2cm,angle=0]{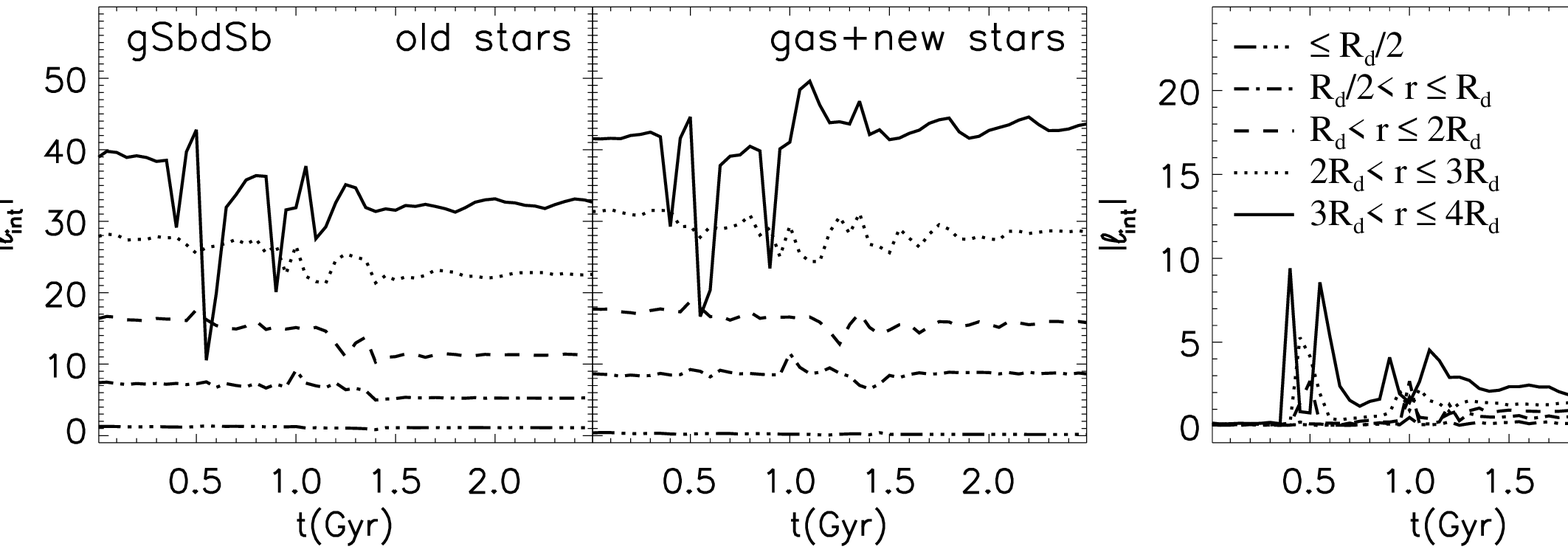}
\vspace{0.5cm}
\caption{Evolution with time of the specific AM, $l$, of a gSb primary galaxy
during the minor merger whose evolution is shown in Fig.\ref{all-3}.
The specific AM of old stars (\emph{left panel}), gas and new stars
(\emph{middle panel}) and dark matter (\emph{right panel}) is shown,
for five different radial zones in the galaxy, see the
legend. $R_{d}$ indicates the disk scale length. The specific AM is in units of
100 kpc km s$^{-1}$.}
\label{amshell}
\end{figure}

During the interaction, the effects of dynamical friction and tidal forces
redistribute the AM in the system: both the primary and the satellite
galaxy acquire part of the orbital AM, and, as already found in Paper I,
the most extended regions experience these changes first. In particular,
the impact of a minor merger on the AM content of the primary galaxy
is shown in Fig.~\ref{amshell}, where the evolution with time of the
specific AM of old stars, gas, new stars and dark matter is shown for
different regions of the galaxy. The figure shows clearly that:

\begin{itemize}
\item the outermost regions of the initially non-rotating dark halo are
the first to acquire part of the orbital AM, already after the first
pericenter passage, at $t=0.5$~Gyr, while inside 3$R_{d}$ (the initial disk
scale length $R_{d}$ is 4.8, 4.8 and 5.2~kpc for S0, Sa and Sb type galaxies,
respectively) tend to increase their AM only during the final phases of
the merger;

\item the AM of the baryonic components is also affected by the
interaction. The old stars experience a slowing down of their rotation
at all radii, especially outside 0.5$R_{d}$. A less pronounced decrease
of the AM is visible in the gas (and new stars) between $R_{d}$ and
3$R_{d}$. For example, in the merger case shown in Fig.~\ref{amshell}
the fractional decrease in the specific AM of  old stars is $\Delta l/l=0.19$
at 2-3$R_d$ whereas $\Delta l/l=0.09$ for gas (and new
stars) in the same region. While the outermost regions
gain AM in the innermost ones the specific AM remains unchanged.

\end{itemize}

\begin{figure}
\centering
\vspace{0cm}\hspace{-0.2cm}
\includegraphics[width=7.5cm,angle=0]{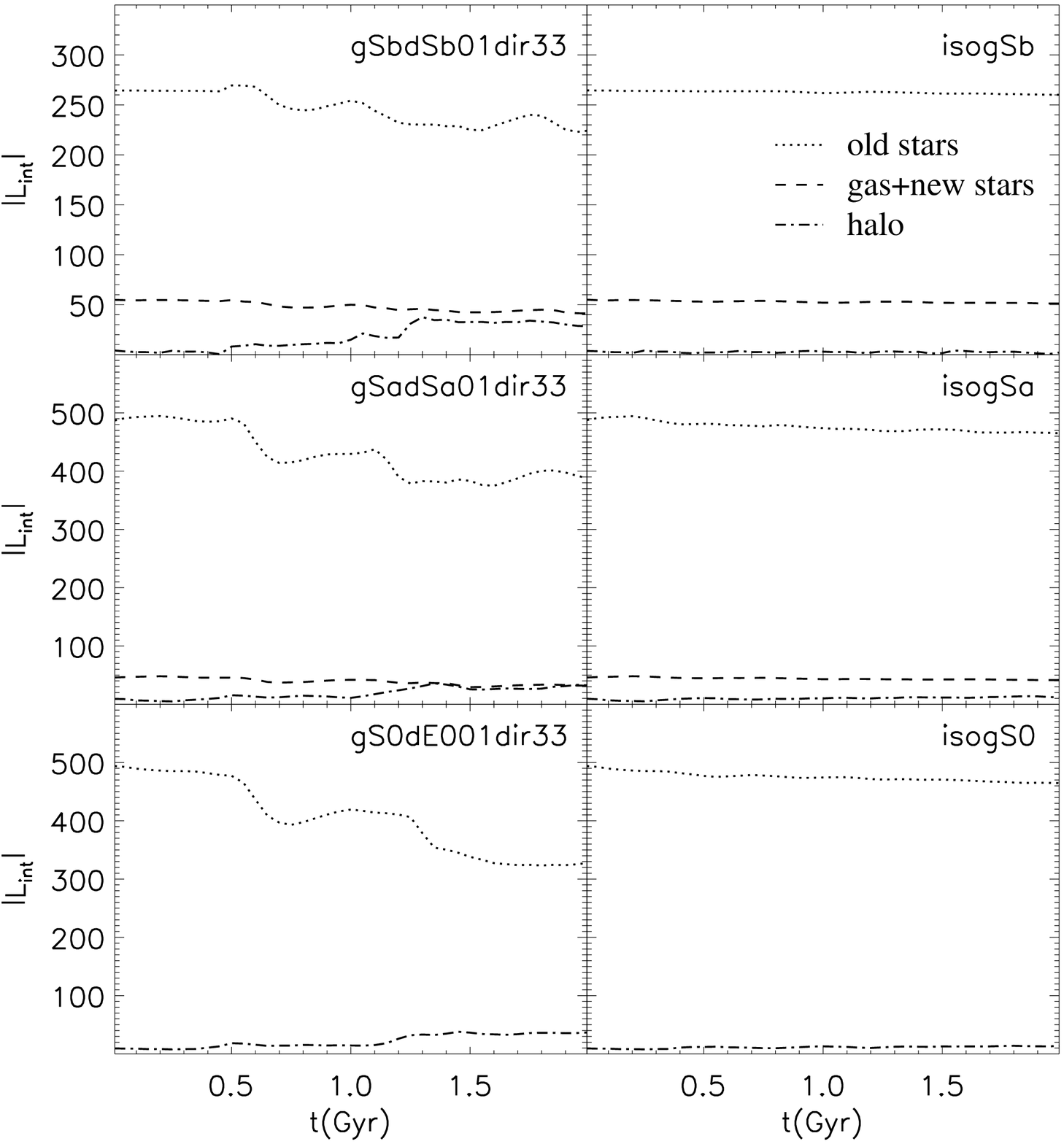}
\vspace{0.7cm}
\caption{\emph{Left panels:} Evolution with time of the total AM of
gas + new stars (dashed line), old stars (dotted line) and the dark matter
halo (dashed-dotted line) in the region at r$<$20 kpc from the center of the
primary galaxy during a minor merger with a satellite galaxy on a direct
orbit. From top to bottom, the gas mass fraction of both primary galaxy and
merging satellite are $20\%$, $10\%$ and 0, respectively. The AM is in
units of 2.2$\times$10$^{11}$M$_{\odot}$ kpc km s$^{-1}$. \emph{Right panels:}
The evolution of the AM of the corresponding components (gas + new stars,
old stars and dark matter halo) of isolated galaxies with the same initial
structure as the primary galaxies in the minor merger simulations.}
\label{totAMisomer}
\end{figure}

The slowing down with time of the AM content of old stars
at a given radius is not only due to disk heating, but also an effect
of the merger process \citep[as has been previously shown
by][]{quillen209}. To demonstrate this effect, in
Fig.~\ref{totAMisomer} we show the total AM of old stars, gas and new stars
(for dissipative mergers), and dark matter of the primary galaxy
for three simulations with different gas fractions in the
primary disk (0, 0.1 and 0.2). In all cases, one can see a
substantial decrease of the total AM of old stars
($\Delta L/L=0.13$ for a gSbdSb direct merger, $0.2$ for a gSadSa merger and
$0.36$ for a dissipationless merger). Also the gas (and new stars)
tend to lose a fraction of their AM. At the same time, the dark
halo of the primary galaxy acquires part of the AM, which is already
apparent after the first passage of the satellite, around
$t=0.5$~Gyr in the simulations. These substantial changes in AM
content of the different components are clearly an effect of the
minor merger itself. This can be seen when comparing the evolution
of the merging galaxies with that of the corresponding galaxies
which were evolved in isolation (right panels of Fig.~\ref{totAMisomer}).
Secular processes, which dominate the evolution of the isolated galaxy
simulations, cause less pronounced changes in the AM, with the decrease
in $\Delta L/L$ of the old stellar disk being only about 20-25\%
of what is estimated for the minor merger simulations.
This demonstrates that disk heating due to secular processes, which
increases the radial stellar velocity dispersions, cannot be the
only mechanism responsible for the slowing down of the disks
observed in the minor merger simulations.

\begin{figure}
\vspace{0.5cm}\hspace{-0.4cm}
\includegraphics[width=9.2cm,angle=0]{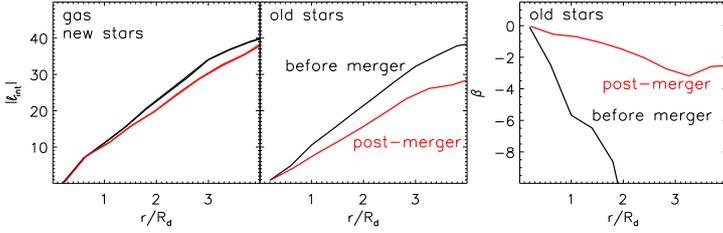}
\vspace{0.5cm}
\caption{\emph{From left to right:} Specific AM as a function of radius
normalized by the disk scale length, $r/R_d$, of the primary galaxy,
for gas and new stars (\emph{left panel}) and old stars (\emph{middle
panel}) during the minor merger simulation whose evolution is shown in
Fig.\ref{all-3}. In each panel the black line shows the initial AM and
the red line shows the final AM. The specific AM is in units of 100 kpc
km s$^{-1}$. \emph{Right panel:} The anisotropy parameter $\beta$ as a
function of $r/R_d$ for old stars in the same merger. Both the initial
(black line) and final (red line) $\beta$ are shown.}
\label{am}
\end{figure}

\begin{figure}
\hspace{0.6cm}
\includegraphics[width=7.5cm,angle=0]{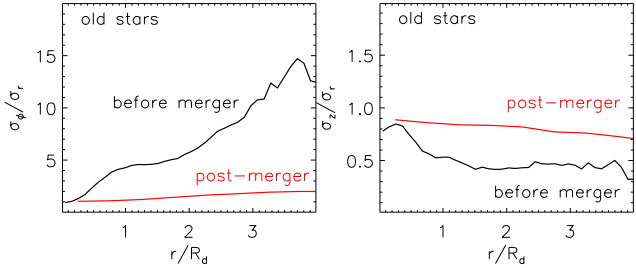}
\vspace{0.7cm}
\caption{\emph{Left panel:} Ratio of tangential and radial velocity
dispersion as a function of scaled radius $r/R_d$ for old stars
before (black line) and after (red line) the minor merger shown in
Fig.\ref{all-3}. \emph{Right panel:} Same as the left panel but for the
ratio of vertical and radial velocity dispersion.}
\label{beta44}
\end{figure}

Figs.~\ref{amshell} and ~\ref{totAMisomer} suggest that it
should be possible to find a difference in the AM content of old and
new stars in the final, post-merger disk, as is indeed the case, see
Fig.~\ref{am}: comparing the specific AM as a function of radius for
gas and new stars (left panel) and old stars (middle panel), one can
see that while the AM of the new stellar component is unchanged
after the merger, that of the old stars decreases at all radii
($\Delta l/l=0.23$ at $R_{d}$). This is due to the fact that, during
the merging process, old stars are heated as shown by the increasing
radial and vertical velocity dispersion (Fig.~\ref{beta44}), which
then leads to a slowing down of the stellar disk, as we have argued
in paper I. However, gas can dissipate its energy, and thus can
preserve mainly tangential motion and keep its AM unaffected,
whereas stars newly formed from this gas retain the AM that was
acquired by the gas. This means that at a given radius the old and
the new stellar populations have a different specific AM and thus
different amounts of rotational support. As discussed in Paper I for
dissipationless mergers, the decrease in the rotation speed of the
old stellar disk of the primary galaxy is accompanied by a change in
the distribution of the types of stellar orbits: the radial
component of the velocity dispersion becomes increasingly important
during the merger, thus increasing the anisotropy parameter,
$\beta=1-\frac{{\sigma_t}^2}{2{\sigma_r}^2}$, from its initially
negative value. This is also the case for dissipative minor mergers 
(Fig.~\ref{am}). \emph{Independent of the amount of gas
present in the primary disk or in the satellite galaxy, minor
mergers always result in a slowing down of the old stellar disk.}
The fractional decrease in the specific AM at $R_{d}$ is $\Delta
l/l=0.23$ for a gSb model, $\Delta l/l=0.17$ for a gSa and $\Delta
l/l=0.29$ for a gS0.

\begin{figure}
\centering
\vspace{1.2cm}\hspace{-8.5cm}
\includegraphics[width=1.2cm,angle=0]{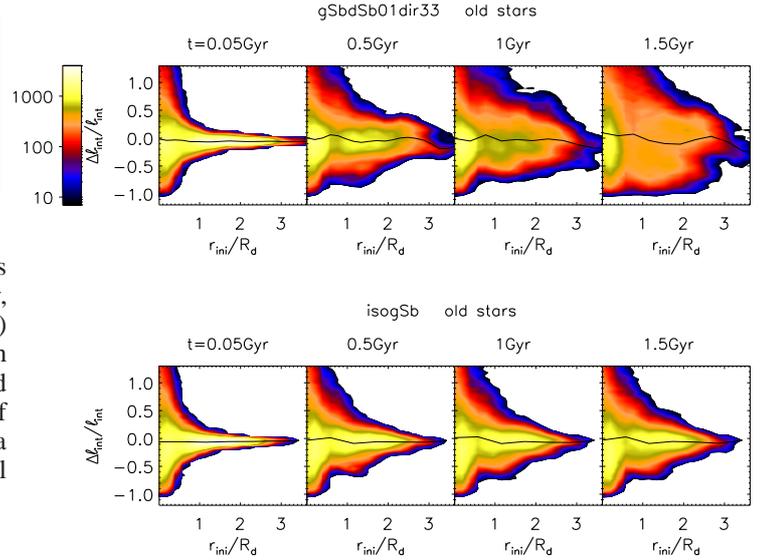}
\vspace{0.7cm}
\caption{\emph{Upper panels}: Relative change of
the specific AM with respect to the initial value, $\Delta
l_{int}/l_{int}=(l_{int}(t)-l_{int}(t=0))/l_{int}(t=0)$, of old stars
of a gSb galaxy as a function of initial radius r$_{\rm ini}$ (scaled
by R$_d$) during a minor merger with a dSb galaxy on a direct orbit
(gSbdSb01dir33). Also shown is the average of $\Delta l_{int}/l_{int}$
as a function of $r_{\rm ini}/R_d$ in each epoch (overlaid solid line).
The vertical bar on the left shows the color scale indicating the number
of old stars at certain $\Delta l_{int}/l_{int}$ values.
\emph{Lower panels:} The distribution of $\Delta l_{int}/l_{int}$ of old
stars of an isolated galaxy with the same initial structure as the
primary galaxy in the minor merger case.}
\label{birthradii}
\end{figure}

Note that the slowing down of the old stellar disk during the merger
is also visible in the time evolution of the AM of those stars that were
initially (at t=0) present in a given radial bin, rather than in that of
stars currently present at a given radius (Fig.~\ref{birthradii}). At
any given initial radius, r$_{ini}$, the variation of the AM is much
stronger when the stellar disk undergoes a minor merger than when it
evolves in isolation. In the minor merger simulations, the average value
of $\Delta l/l$ decreases with time, especially for stars whose initial
radii exceed $R_d$. The largest changes take place in the
outer disk where the effect of the perturbations due to the satellite is
the greatest \citep[as already has been shown by][]{quillen209, bird211}.

\subsubsection{The impact of multiple mergers}

\begin{figure}
\hspace{0.4cm}
\begin{minipage}{0.2\textwidth}
\centering
\includegraphics[width=3.8cm,angle=0]{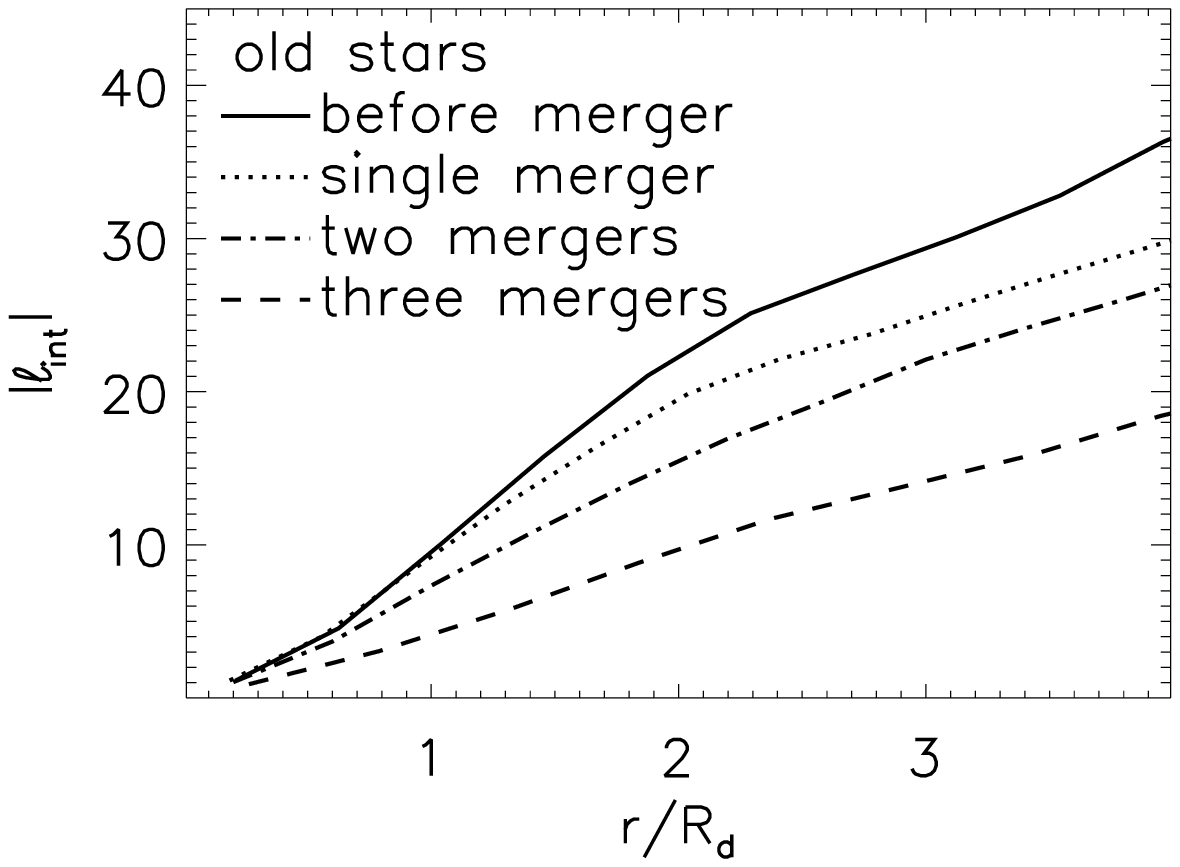}
 \end{minipage} \hspace{0.8cm}
 \begin{minipage}{0.2\textwidth}
\centering
\includegraphics[width=3.8cm,angle=0]{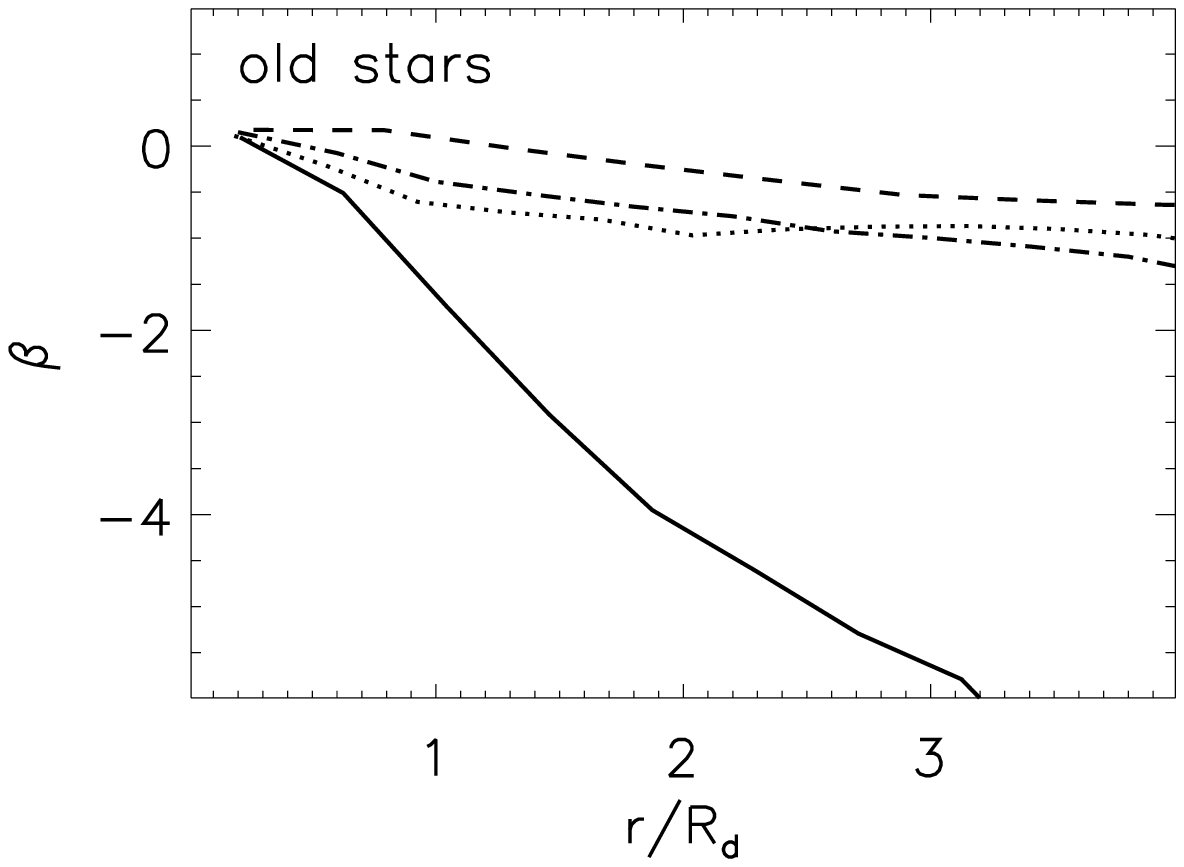}
 \end{minipage} \vspace{0.7cm}
\caption{\emph{Left panel:} Specific AM as a function of scaled radius
$r/R_d$ for the old stars of a gSa galaxy undergoing three consecutive
minor mergers. The specific AM is shown after the first (dotted line),
the second (dashed-dotted line), and the third (dashed line) merger, and the initial
AM is shown for comparison (solid line). The specific AM is in units of
100~kpc km s$^{-1}$. \emph{Right panel:} Anisotropy parameter, $\beta$,
as a function of scaled radius, for the old stars of the gSa galaxy
whose specific AM is shown in the left panel.}
\label{multi-am-beta}
\end{figure}

\begin{figure}
\vspace{0cm}\hspace{0.1cm}
\begin{minipage}{0.2\textwidth}
\centering
\includegraphics[width=3.8cm,angle=0]{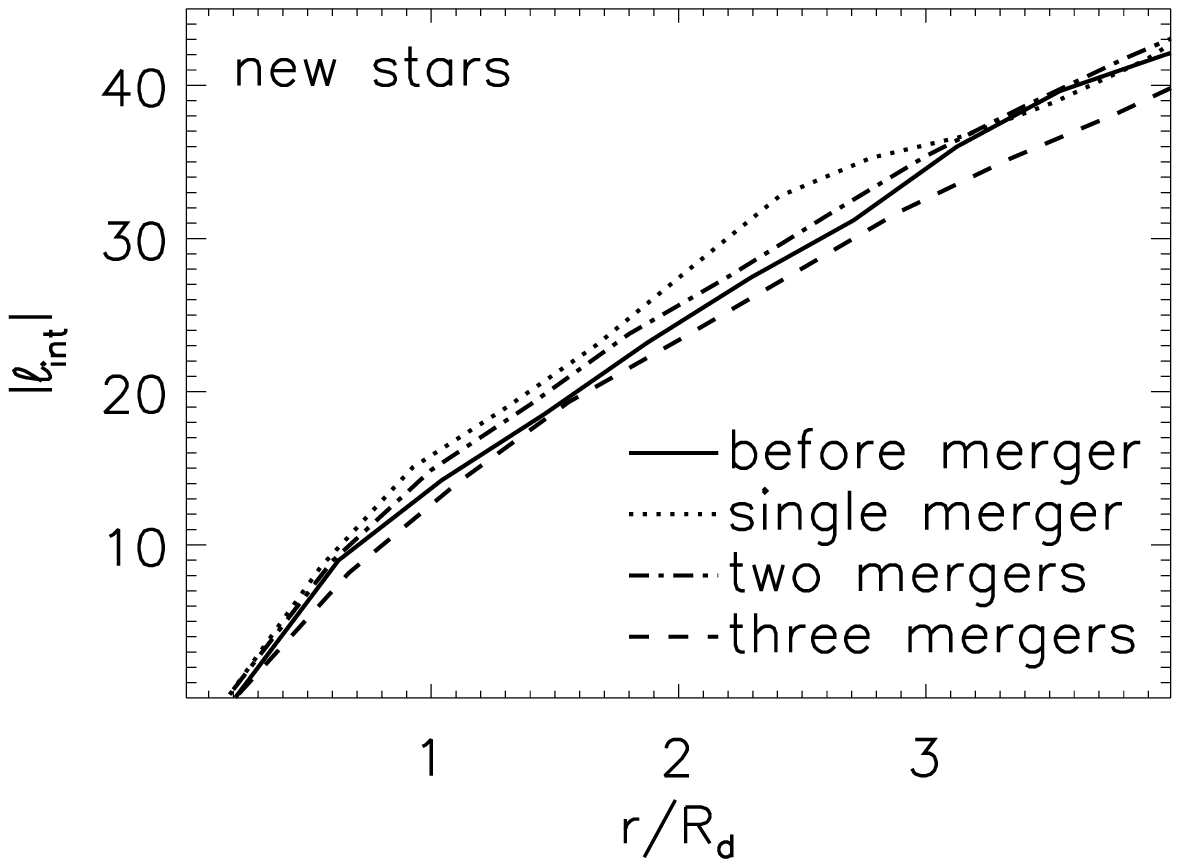}
 \end{minipage} \hspace{1.2cm}
 \begin{minipage}{0.2\textwidth}
\centering
\includegraphics[width=3.8cm,angle=0]{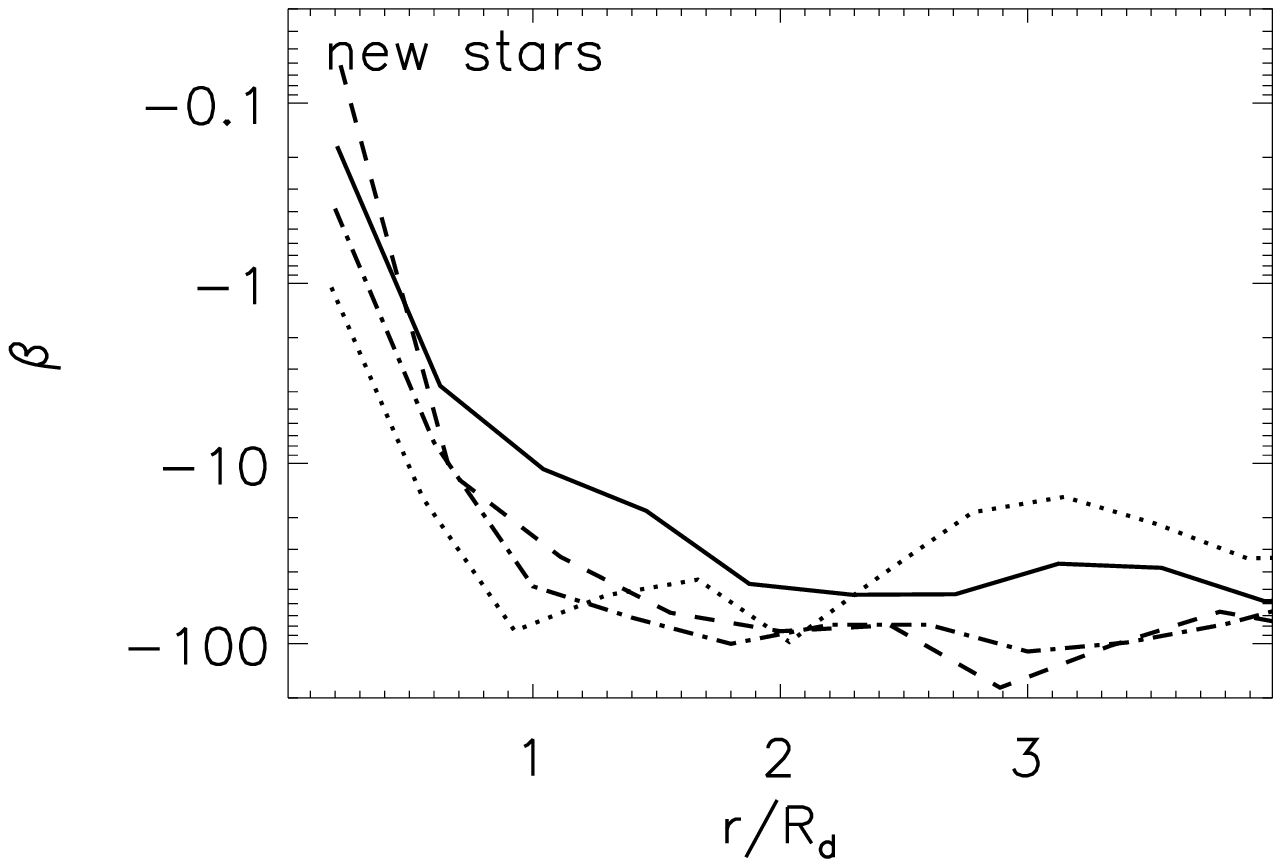}
 \end{minipage} \vspace{0.7cm}
\caption{Same as Fig.~\ref{multi-am-beta} but for newly formed stars in the gSa galaxy.}
\label{multi-am-beta2}
\end{figure}

If a single minor merger causes the slowing down of the old stellar
component of a galaxy disk, what would be the effect of a subsequent
minor merger on it? To answer this question, we ran ten simulations
in which the primary galaxy consecutively accretes two ten times
less massive satellites over a period of 3-5~Gyr. The second minor
merger produces a further decrease in the specific AM of the old
stellar population (Fig.~\ref{multi-am-beta}).  After the first
merger, the specific AM decreases at all radii, accompanied by an
increase of the $\beta$ parameter. In the second merger event, the
AM content of the old disk stellar component is further decreased
and this is accompanied by an increase in the velocity anisotropy
parameter $\beta$ at all radii in the disk. Furthermore, we ran one
simulation in which the primary galaxy consecutively accretes three
ten times less massive satellites over a period of 5~Gyr. After
having accreted two satellites the stellar disk of the primary
galaxy still reacted dynamically to the third satellite accretion --
its specific AM decreased further, its $beta$ parameter increased
and the slowing down of the old stellar disk did not saturate but
increased (see Fig.~\ref{multi-am-beta}). While old stars slow
down with time, whereas stars formed during the repeated mergers
tend to show little change in their AM (Fig.~\ref{multi-am-beta2}),
the difference in AM content between stars born at different times
increases during repeated mergers with low mass satellite galaxies.
This is only partially due to the hybrid method used to implement
star formation in our simulations, in which the new stars are
continuously formed from dissipative gas which keeps the AM mostly
unchanged. Even though secular disk heating processes also have an
effect on heating newly formed stars, this effect is significantly
lower than that induced by minor mergers, as already shown in
Fig.~\ref{totAMisomer}.

\subsection{Rotational lag}\label{rotlag}

\begin{figure*}
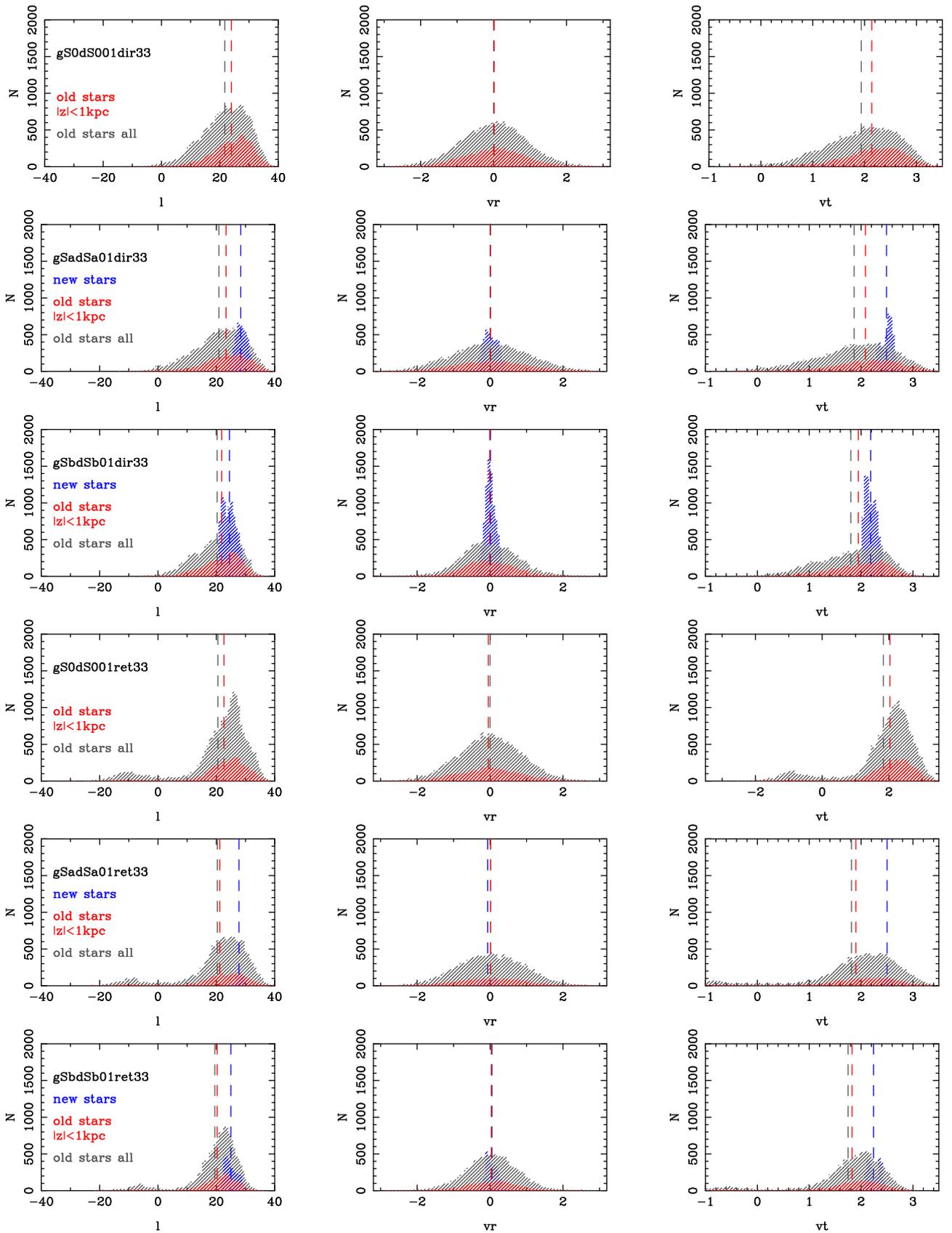

\vspace{0cm}\hspace{0.0cm}
\begin{minipage}{1.\textwidth}
\centering
\includegraphics[width=3.7cm,angle=270]{16502_11.ps}
 \end{minipage} \hspace{0.cm}
 \begin{minipage}{1.\textwidth}
\centering
\includegraphics[width=3.7cm,angle=270]{16502_12.ps}
 \end{minipage} \hspace{0.cm}
 \begin{minipage}{1.\textwidth}
\centering
\includegraphics[width=3.7cm,angle=270]{16502_13.ps}
 \end{minipage} \hspace{0.cm}
 \begin{minipage}{1.\textwidth}
\centering
\includegraphics[width=3.7cm,angle=270]{16502_14.ps}
 \end{minipage} \hspace{0.cm}
 \begin{minipage}{1.\textwidth}
\centering
\includegraphics[width=3.7cm,angle=270]{16502_15.ps}
 \end{minipage} \hspace{0.cm}
\begin{minipage}{1.\textwidth}
\centering
\includegraphics[width=3.7cm,angle=270]{16502_16.ps}
\end{minipage}  \hspace{0.cm}
\vspace{0.3cm}
\caption{Histograms of the specific AM, $l$ (in units of 100~kpc km
s$^{-1}$), and the radial ($v_r$) and tangential ($v_t$) velocities
(in units of 100~km s$^{-1}$) of stars between radii of 2-3$R_{d}$
in a number of minor merger remnants.  The top three rows show
direct minor mergers, with disk gas mass fractions increasing from 0
to 0.2. Indicated in each panel are the entire old stellar
population (gray), new stars (blue) and old stars with $\mid
z\mid\le$1~kpc (red). The vertical dashed lines show the average
values of $l$, $v_r$ and $v_t$ for each of these three components.
The three lowest rows are similar, but for retrograde mergers.}
\label{histo}
\end{figure*}

\begin{figure*}
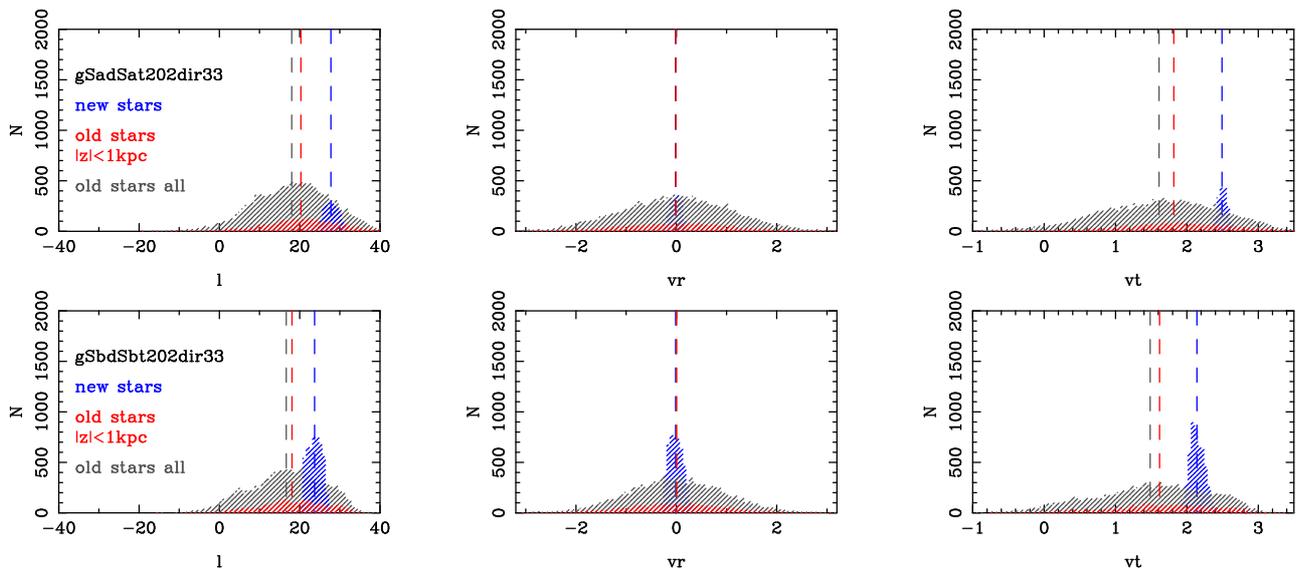

\begin{minipage}{1.\textwidth}
\centering
\includegraphics[width=3.7cm,angle=270]{16502_17.ps}
\end{minipage} \hspace{0.cm}
\begin{minipage}{1.\textwidth}
\centering
\includegraphics[width=3.7cm,angle=270]{16502_18.ps}
\end{minipage}\vspace{0.2cm}
\caption{Same as Fig.~\ref{histo}, but for two consecutive minor mergers.}
\label{histo2}
\end{figure*}

\begin{figure*}
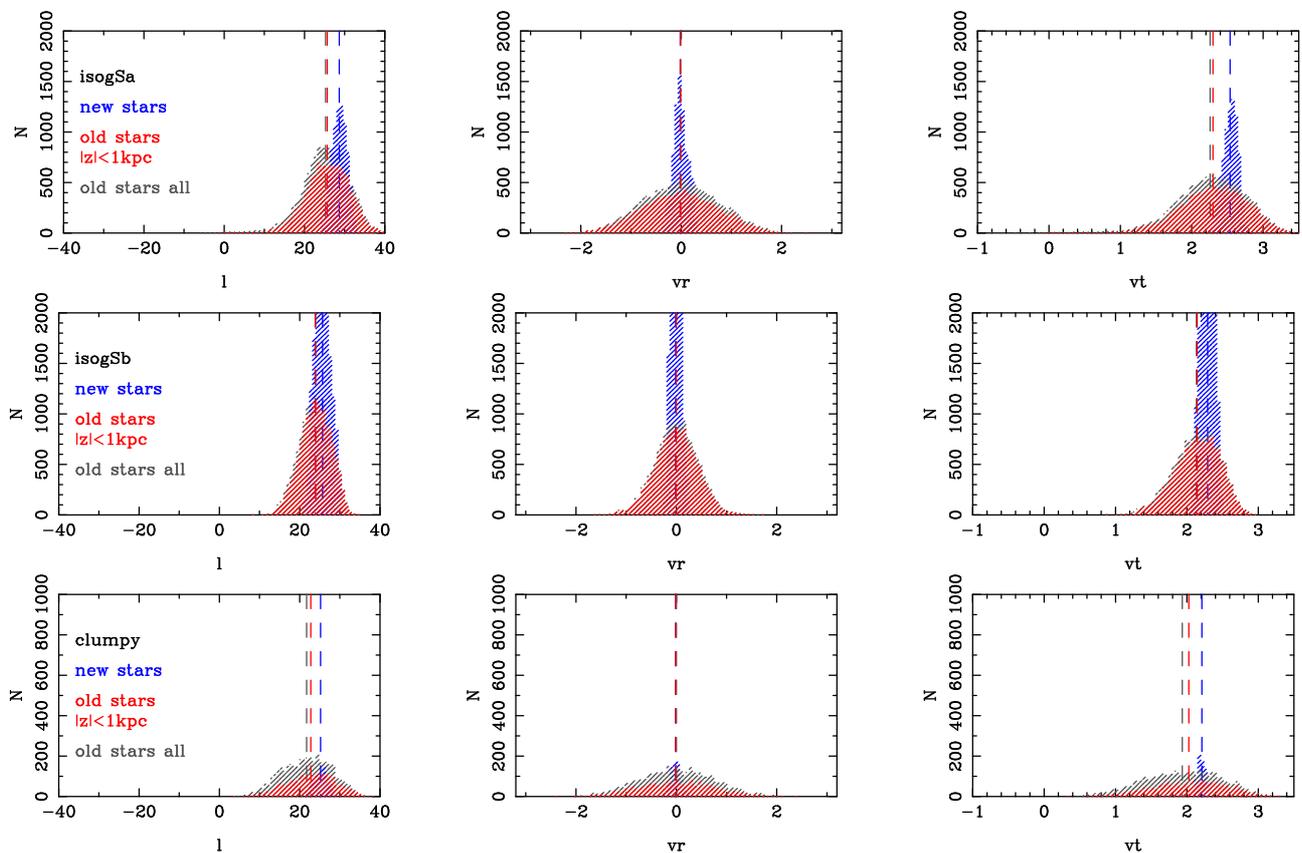

\vspace{0.cm}\hspace{0.cm}
\begin{minipage}{1.\textwidth}
\centering
\includegraphics[width=3.7cm,angle=270]{16502_19.ps}
\end{minipage} \hspace{0.cm}
\begin{minipage}{1.\textwidth}
\centering
\includegraphics[width=3.7cm,angle=270]{16502_20.ps}
\end{minipage} \hspace{0.cm}
\begin{minipage}{1.\textwidth}
\centering
\includegraphics[width=3.7cm,angle=270]{16502_21.ps}
\end{minipage}\vspace{0.2cm}
\caption{Same as Fig.~\ref{histo}, but for galaxies that were evolved
in isolation for 3~Gyr: a gSa (upper panels) with a gas mass fraction $f_{\rm gas}=0.1$ and a
gSb (middle panels) with $f_{\rm gas}=0.2$, as well as a galaxy with a much
higher gas fraction of 0.5 which went through an unstable clumpy phase
(bottom panels). In the latter case the parameters are also shown after
3~Gyr, when the clumps have dissolved through tidal effects or spiraled into
the galaxy center through dynamical friction and interaction.}
\label{histo-iso}
\end{figure*}

As the slowing down is only affecting the old stellar population, we
expect to see an increasing difference between the AM content of the
old and new stellar components as mergers proceed. This is indeed
the case, as shown in Figs.~\ref{histo} and \ref{histo2}, where the
distribution of the specific AM, $l$, and radial and tangential
velocities ($v_r$ and $v_t$, respectively) of stars between radii of
2 and 3$R_d$ are shown after a single minor merger and after two
consecutive mergers. We chose this radial region as it should be
representative of the solar neighborhood.  To compare the effects of
consecutive mergers to those of secular evolution, we show in Fig.~\ref{histo-iso} 
the same distributions for three galaxy simulations that do not 
undergo a merger but are evolved in isolation: 
for gSa and gSb model galaxies, and for a gas-rich, 
unstable clumpy disk.  In this plot we show the newly formed stars
as well as the entire old star population and the contribution of
old stars at heights $\mid z\mid\le$1~kpc, in order to distinguish
old stars distributed in a thin disk from the total. We emphasize
that this is a very simple way to distinguish between a thin,
dynamically cold stellar component and a thick and hotter stellar
component, which is based on the analysis of the vertical properties
of thick disks formed in minor mergers \citep{quDM211b}.

The results of this study show that:

\begin{itemize}
\item After a single merger the thin stellar disk (i.e., at $\mid
z\mid\le$1~kpc) consists of two different stellar components, a young
and an old population, the latter with a lower angular momentum than
the former. In the region between 2 and 3$R_{d}$, for a gSbdSb direct
merger, where both galaxies initially have a gas fraction of 20$\%$,
the old stars in the thin disk have an average specific AM which is
about 10$\%$ lower than the new stars;

\item If the whole old stellar population is considered, its average
specific AM is about 7$\%$ lower than that of old stars at $\mid
z\mid\le$1~kpc, indicating that the stars in the thick old disk ($\mid
z\mid \ge$1~kpc) contribute to  decrease the specific AM  and are thus
less rotationally supported than the old stars in the thin disk;

\item This difference in specific AM content of the stellar populations
is reflected in different tangential velocities of their stars in the
disk, with the tangential support of stellar orbits decreasing from the
thin new disk, through the thin old disk, to the thick old disk;

\item For the same gas fraction in the progenitor disks there is a
difference between direct and retrograde mergers, in the sense that
the specific AM content of old stars is lower if the satellite orbit
is retrograde. This lower AM content can be explained by the presence
of counter-rotating material (negative $l$\ ) which is found in all
retrograde encounters analyzed (see \S~\ref{counter} for a detailed
discussion);

\item If one defines the rotational lag of old stars as
$v_{lag}=v_{*}-v_{new stars}$, where $v_{*}$ is the tangential velocity
of stars in the thin or thick old disk, and $v_{new stars}$ is that of
new stars, one can see that it depends on the amount of gas present in
the progenitor disks: for old thin and old thick disk stars, \emph{the
higher the gas fraction in the progenitor disk, the lower the rotational
lag in the remnant.}

\end{itemize}

We refer the reader to Table~\ref{lagtable} for a complete summary of
the results found in Figs.~\ref{histo}, \ref{histo2} and \ref{histo-iso}.

The difference between the specific AM content of old and new stars
becomes still more pronounced after a second, consecutive minor merger,
leading to a further increase in the rotational lag of both the old
thin and thick disk stars, as shown in Fig.~\ref{histo2}. Given that
minor mergers have a cumulative impact on heating and slowing down
stellar disks (see also Fig.~\ref{multi-am-beta}), two mergers of mass
ratio 1:10 are expected to result in a difference between the specific
AM content of old and new stars that is comparable with that of one
merger of mass ratio 1:5. In other words, more massive satellites can
have a larger impact on slowing down the stellar disk and produce a
larger rotational lag between the old and new stellar populations than
less massive satellites \citep{quDM210a}. The substantial difference
between the specific AM content of the thin and thick disk stellar
populations cannot be due to secular evolution processes alone, which
produce a considerably smaller variation in both $l$ and $v_t$ in 3~Gyr
of evolution (Fig.~\ref{histo-iso}). Note also that secular evolution
produces a narrower distribution of $l$, $v_r$ and $v_t$, thus suggesting
that also mixing and radial migration due to secular processes may be less
effective than in minor mergers. Whereas scattering of stars by massive
clumps can induce a rotational lag (in old thin and thick disk stars),
the overall impact of scattering by mass concentrations is (moderately)
lower than that produced by a direct 1:10 merger on a galaxy with an
initial gas fraction of 20$\%$.

\begin{table*}
\caption[]{Kinematics of stars in the post-merger thin and 
thick old thick disks and of the new stars in disk galaxies heated by
different physical processes. For old stars in the thin and the thick disks
the rotational lag is defined with respect to the tangential
velocity $v_t$ of the new stars.}
\label{lagtable} \centering
\begin{tabular}{cccccccccc}
\hline\hline
\multicolumn{2}{c}{}&
\multicolumn{3}{c}{old stars $\mid z\mid\leq$1~kpc}&
\multicolumn{3}{c}{old stars $\mid z\mid >$1~kpc}&
\multicolumn{2}{c}{new stars}
\\\hline
heating mechanism & $f_{\rm gas}$ & mass & $v_t$ & lag &  mass fraction & $v_t$ & lag&  mass fraction & $v_t$  \\
& &  [stellar mass] & [km s$^{-1}$] & [km s$^{-1}$] & [stellar mass] & [km s$^{-1}$]&  [km s$^{-1}$]&  [stellar mass] & [km s$^{-1}$] \\
\hline
1:10 merger (direct)& 0. & 0.38  & 214. & n.d. & 0.62 & 181. & n.d. & -- & --  \\
1:10 merger (retrograde)&0. & 0.28 & 203.  & n.d. & 0.72 & 176. & n.d. & -- & --  \\

1:10 merger (direct) & 0.1 & 0.34 & 208. & 42. & 0.65 & 175. & 75. & 0.01 & 250. \\
1:10 merger (retrograde) & 0.1 & 0.23 & 190. &  60. & 0.77  & 179. & 71. & 0.001 & 250. \\

1:10 merger (direct) & 0.2 & 0.4 & 195. & 24. & 0.58 & 170. & 49. & 0.015 & 219.\\
1:10 merger (retrograde) & 0.2 & 0.24 & 183. & 41. & 0.75  & 172. & 52. & 0.004 & 224.  \\

2x(1:10) merger & 0.1 & 0.22 & 182. & 68. & 0.77 & 155. & 95. & 0.01& 250. \\
2x(1:10) merger & 0.2 & 0.25 & 162. & 52. & 0.74 & 144. & 70. & 0.01& 214.\\
secular & 0.1 & 0.79 & 230. & 24. & 0.19 & 210. & 44. & 0.015 & 254. \\
secular & 0.2 & 0.91 & 214. & 16. & 0.07 & 200. & 30. & 0.019 & 230.\\
clumpy & 0.5 & 0.53 & 203. & 17. & 0.49 & 183. & 37. & 0.016& 220.\\
\hline
\end{tabular}
\end{table*}

\subsection{Accreted and heated disk stars}\label{discriminate}

\subsubsection{Contribution to the rotational lag}\label{discriminate1}

\begin{figure*}
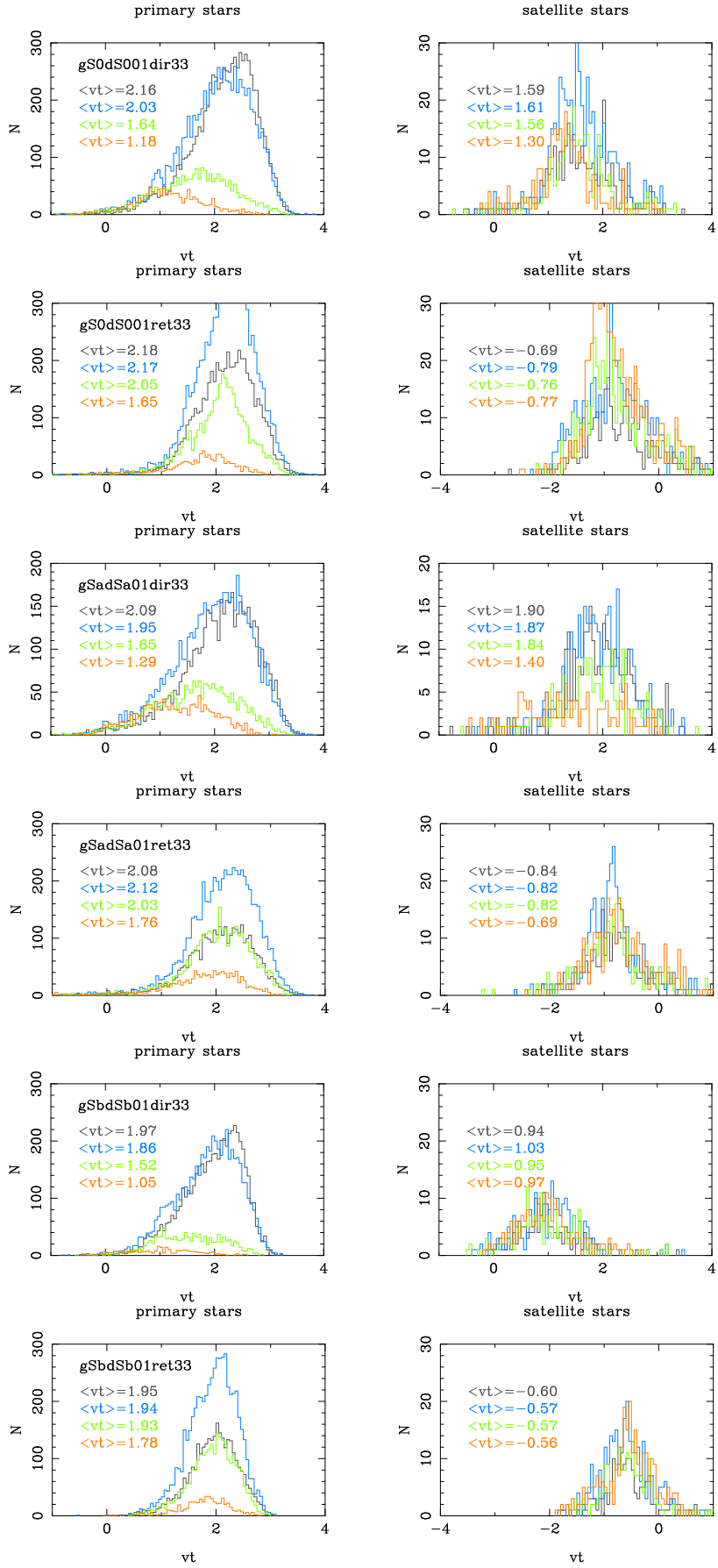

 \begin{minipage}{1.\textwidth}
\centering
\includegraphics[width=3.9cm,angle=270]{16502_22.ps}
 \end{minipage} \hspace{0.cm}
 \begin{minipage}{1.\textwidth}
\centering
\includegraphics[width=3.9cm,angle=270]{16502_23.ps}
 \end{minipage} \hspace{0.cm}
 \begin{minipage}{1.\textwidth}
\centering
\includegraphics[width=3.9cm,angle=270]{16502_24.ps}
 \end{minipage} \hspace{0.cm}
 \begin{minipage}{1.\textwidth}
\centering
\includegraphics[width=3.9cm,angle=270]{16502_25.ps}
 \end{minipage} \hspace{0.cm}
 \begin{minipage}{1.\textwidth}
\centering
\includegraphics[width=3.9cm,angle=270]{16502_26.ps}
 \end{minipage} \hspace{0.cm}
 \begin{minipage}{1.\textwidth}
\centering
\includegraphics[width=3.9cm,angle=270]{16502_27.ps}
\end{minipage}\vspace{0.2cm}
\caption{Distribution of tangential velocities $v_t$ (in units of
100 km s$^{-1}$) of old stars for six minor mergers. For each
encounter a pair of panels is shown, with stars originally in the
primary disk (left) and those accreted from the satellite (right).
In each panel the histograms correspond to different heights above
and below the galaxy midplane: $\mid z\mid\le$1~kpc (black), 1$<\mid
z\mid\le$3~kpc (blue), 3$<\mid z\mid\le$5~kpc (green) and 5$<\mid
z\mid\le$10~kpc (orange). The average values of $v_t$ in the
different regions is also indicated.} \label{transvel}
\end{figure*}

In the previous section we saw that old stars in the thin and thick
disk of a minor merger remnant lag with respect to the new stars,
and that this lag is higher for stars further from the galaxy
midplane ($\mid z\mid\ge$1~kpc). But what is this rotational lag due
to -- is it mostly associated to stars originally in the satellite
or in the primary disk, and how does it depend on the orbital
parameters?

\begin{figure}
\centering
\includegraphics[width=4.5cm,angle=270]{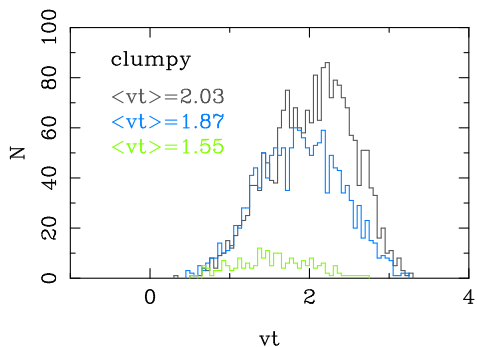}
\vspace{0.2cm} \caption{Same as Fig.~\ref{transvel} but for an
isolated, very gas-rich galaxy ($f_{\rm gas}=0.5$), after an
unstable, clumpy phase. Note that in this case the fourth and
highest region is missing, due to a lack of stars above/below 5~kpc
from the galaxy midplane.} \label{transvel_clumpy}
\end{figure}

To answer these questions, we analyzed the tangential velocity $v_t$
of old stars at radii between 2 and 3$R_d$ in four different regions
above the disk midplane: $\mid z\mid\le$1~kpc, 1$<\mid
z\mid\le$3~kpc, 3$<\mid z\mid\le$5~kpc and 5$<\mid z\mid\le$10~kpc.
Results from some representative mergers are shown in
Fig.~\ref{transvel}, where for each encounter we distinguished the
primary stars from those originally in the satellite.  This figure
shows some interesting trends. First of all, there is a clear
difference between direct and retrograde mergers. While in both
cases the average $v_t$ of old stars is lower than that of new stars
(Table~\ref{lagtable}), the behavior of $v_t$ as a function of
height z is significantly different. Direct mergers produce disks
with a tangential velocity $v_t$ which decreases with height,
meaning that the rotational lag increases with z, whereas in
retrograde mergers $v_t$ is constant with height up to z$\sim$5~kpc,
and decreases only at greater heights. The satellite stars, which in
each region constitute only a small percentage of the total stellar
content, do not contribute significantly to the lag. Moreover, if
one compares the tangential velocities of satellite and primary
stars for direct orbits, one can find a variety of behaviors in the
regions analyzed: in some cases the tangential velocities of
satellite stars are always smaller than those of stars from the
primary (e.g. for orbit ``gS0dS001dir33''), while in other cases the
values are comparable, and satellite stars show even higher
velocities than primary stars in the outer regions (e.g. orbit
``gSadSa01dir33'').  In all the cases, however, the strongest
variations in $v_t$ as a function of z are associated with stars
originally in the primary rather than from the satellite. For
retrograde orbits, the tangential velocities of satellite stars show
only small variations with increasing z, as is the case for primary
stars. We note also that a stellar thick disk formed in an unstable
clumpy galaxy is characterized by tangential velocities whose
variation with z is very similar to that produced in a direct
encounter with $f_{\rm gas}=0.2$ (compare, e.g.
Fig.~\ref{transvel_clumpy} with orbit ``gSbdSb01dir33'' in
Fig.~\ref{transvel}).

\subsubsection{How to distinguish satellite stars}\label{discriminate2}

\begin{figure*}
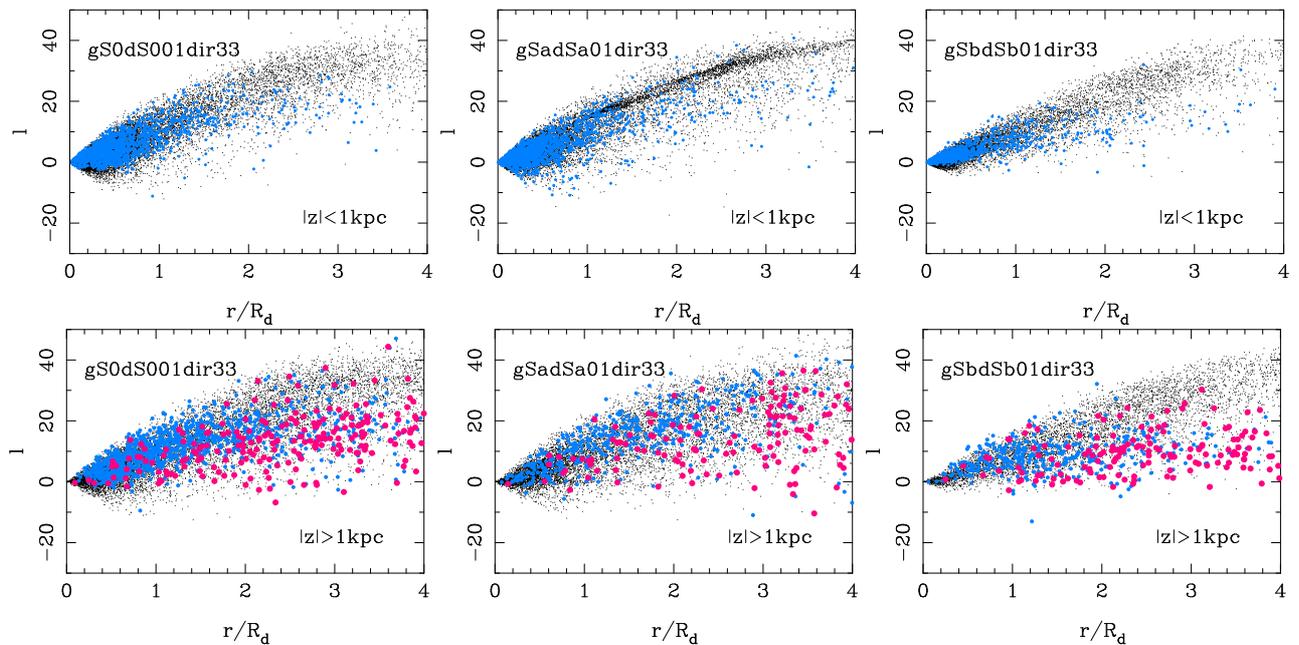

\centering
\includegraphics[width=4.2cm,angle=270]{16502_29.ps}
\includegraphics[width=4.2cm,angle=270]{16502_30.ps}
\includegraphics[width=4.2cm,angle=270]{16502_31.ps}
\includegraphics[width=4.2cm,angle=270]{16502_32.ps}
\includegraphics[width=4.2cm,angle=270]{16502_33.ps}
\includegraphics[width=4.2cm,angle=270]{16502_34.ps}
\vspace{0.2cm} \caption{Specific AM, $l$, as a function of scaled
radius for old stars in three direct merger remnants. For each
encounter a pair of plots is shown, where stars are selected on
their heights above (or below) the galaxy midplane: $\mid
z\mid\le$1~kpc (upper panels) and $\mid z\mid >$1~kpc (lower
panels). In each panel stars originally in the primary galaxy are
shown as black points, stars originally in the satellite as blue
points, and satellite stars with $\mid z\mid >$5~kpc as red points.
The specific AM is in units of 100 kpc km s$^{-1}$.} \label{AMsat1}
\end{figure*}

\begin{figure*}
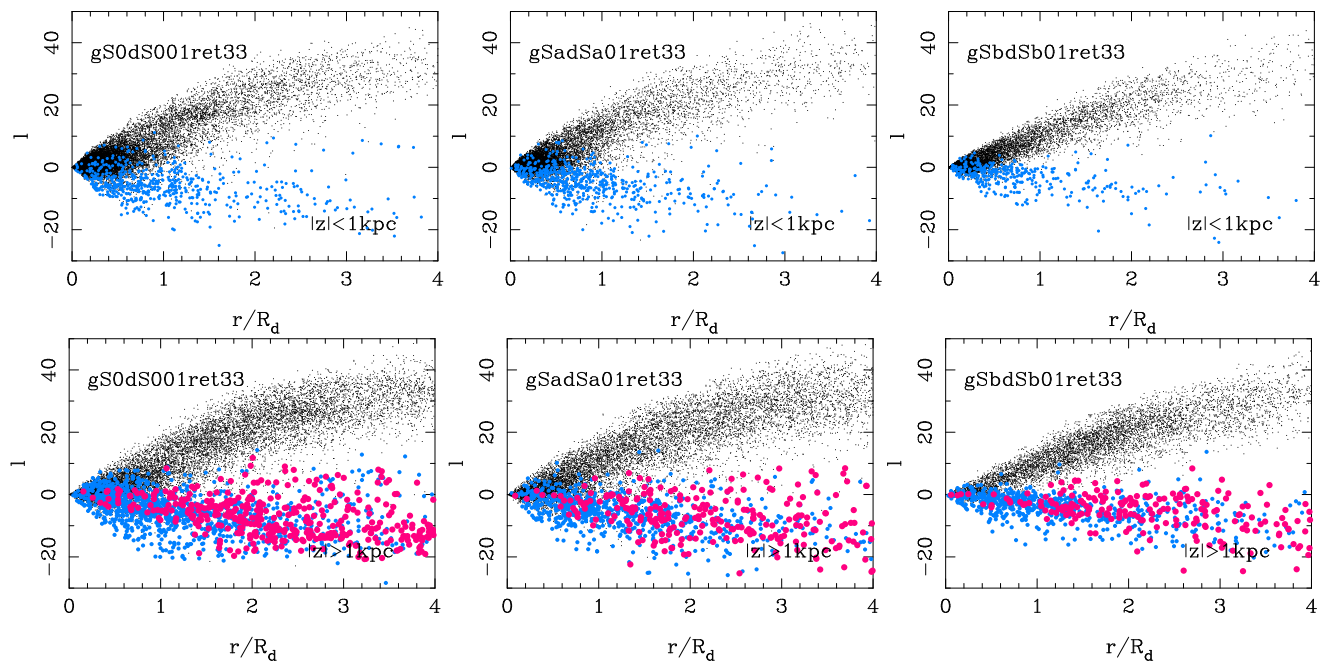

\centering
\includegraphics[width=4.3cm,angle=270]{16502_35.ps}
\includegraphics[width=4.3cm,angle=270]{16502_36.ps}
\includegraphics[width=4.3cm,angle=270]{16502_37.ps}
\includegraphics[width=4.3cm,angle=270]{16502_38.ps}
\includegraphics[width=4.3cm,angle=270]{16502_39.ps}
\includegraphics[width=4.3cm,angle=270]{16502_40.ps}
\vspace{0.2cm}
\caption{Same as Fig.~\ref{AMsat1}, but for three retrograde mergers.}
\label{AMsat2}
\end{figure*}

\begin{figure*}
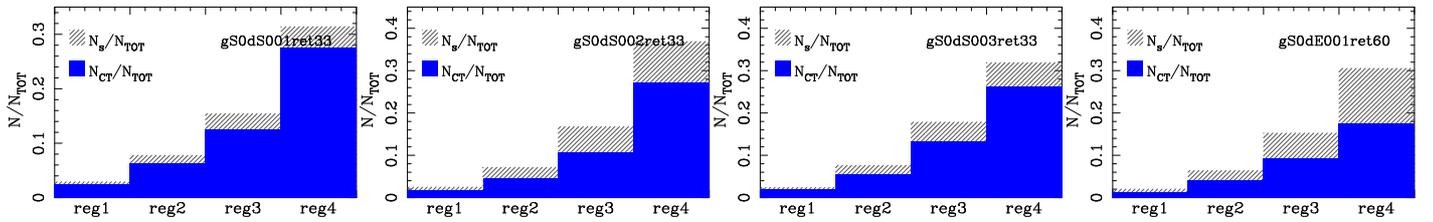

\vspace{0cm}\hspace{-0.2cm}
 \begin{minipage}{0.2\textwidth}
\centering
\includegraphics[width=2.8cm,angle=270]{16502_41.ps}
 \end{minipage} \hspace{0.8cm}
 \begin{minipage}{0.2\textwidth}
\centering
\includegraphics[width=2.8cm,angle=270]{16502_42.ps}
 \end{minipage} \hspace{0.8cm}
 \begin{minipage}{0.2\textwidth}
\centering
\includegraphics[width=2.8cm,angle=270]{16502_43.ps}
 \end{minipage} \hspace{0.8cm}
 \begin{minipage}{0.2\textwidth}
\centering
\includegraphics[width=2.8cm,angle=270]{16502_44.ps}
 \end{minipage} \vspace{0.2cm}
\caption{Fraction of counter-rotating stars (shown in blue) for four
dissipationless retrograde mergers. Stars are divided into four
regions, according to their height above (or below) the galaxy
midplane: (1) at $\mid z\mid\le$1~kpc, (2) at 1~kpc$\le\mid z\mid\le
$3~kpc, (3) at 3~kpc$\le\mid z\mid\le$5~kpc and (4) at 5~kpc$\le\mid
z\mid\le$10~kpc. For each region the fraction of satellite stars to
the total number of stars is also shown (gray).}
\label{counter_nogas}
\end{figure*}

\citet{villalobosH209} recently proposed an interesting method to
use the vertical AM to discriminate between stars formed in the
primary thin disk and then heated by a minor merger from those
originally in the satellite and then accreted. Their simulations
showed (see their Fig.6) that satellite stars in the height zone
$\mid z\mid\le$1~kpc have a typical AM content which is constant
with radius, while stars from the primary disk tend to have an AM
$l\propto r$. Their simulations are gas-free and involve satellites
more massive than the ones studied in this paper.  Because of these
differences we decided to reinvestigate the validity of their
method, exploring if the result is similar for larger mass ratios
like ours (1:10) and for other satellite and primary galaxy
morphologies. Our results for the specific AM of stars originally in
the primary disk and in the satellite galaxy are shown in
Figs.~\ref{AMsat1} and ~\ref{AMsat2}. We divided the stars into two
regions, one at $\mid z\mid\le$1~kpc and another at $\mid
z\mid\ge$1~kpc. The general result confirms that the difference
between the specific AM of satellite stars and primary disk stars is
significant in the outer disk regions: typically the difference is
20$\%$ at radii of 1.5-2$R_d$.  But at these radii the fraction of
satellite stars with $\mid z\mid\le$1~kpc is only a few percent of
the total number of stars. This means that it can be extremely
difficult to detect these stars in an observational sample.  We
find, however, that the fraction of satellite stars is five to ten
times larger at greater heights: at $\mid z\mid\ge$1~kpc, outside a
radius of 1.5-2$R_d$ the specific AM content of accreted stars can
be significantly different from that of primary stars, and their
fraction sufficiently higher (around 10$\%$) to make the
observational detection of this signature more likely. Note however
that generally the satellite stars with the lowest AM content are
found very far from the galaxy midplane ($\mid z\mid\ge$5~kpc). We
want to point out that there are also cases in which there is no
discernible difference in the specific AM of satellite and primary
disk stars at any radius, even at the largest radii plotted in
Figs.~\ref{AMsat1} and \ref{AMsat2}). This is in agreement with
what was found in Fig.~\ref{transvel}, where primary and satellite
stars show remarkably similar tangential velocities.

Obviously, the specific AM becomes a better discriminant if satellite
orbits counter-rotate with respect to the primary disk -- a clear sign
of an external origin of stars. We find counter-rotation in all merger
remnants which formed from satellites on retrograde orbits (some examples
are shown in Fig.~\ref{AMsat2}).  We will discuss the main properties
of counter-rotating stars further in the next section.

\begin{figure*}
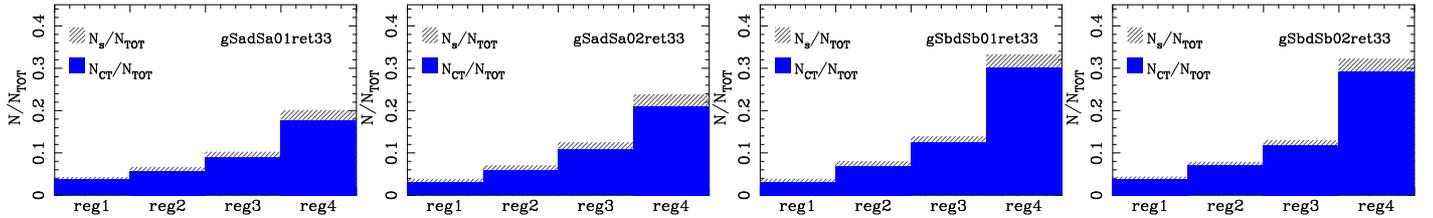

\vspace{0.cm}\hspace{-0.2cm}
 \begin{minipage}{0.2\textwidth}
\centering
\includegraphics[width=2.8cm,angle=270]{16502_45.ps}
 \end{minipage} \hspace{0.8cm}
 \begin{minipage}{0.2\textwidth}
\centering
\includegraphics[width=2.8cm,angle=270]{16502_46.ps}
 \end{minipage} \hspace{0.8cm}
 \begin{minipage}{0.2\textwidth}
\centering
\includegraphics[width=2.8cm,angle=270]{16502_47.ps}
 \end{minipage} \hspace{0.8cm}
 \begin{minipage}{0.2\textwidth}
\centering
\includegraphics[width=2.8cm,angle=270]{16502_48.ps}
 \end{minipage} \vspace{0.2cm}
\caption{Same as Fig.\ref{counter_nogas}, but for four dissipative
retrograde mergers.} \label{counter_gas}
\end{figure*}

\subsection{Counter-rotation}\label{counter}

Counter-rotating stellar disks are observed in external galaxies
\citep[see ][]{yoachimD205} and counter-rotation is also found in
the Milky Way halo \citep[see ][]{carollo207}. The presence of stars
(in the thick disk or in the halo) rotating in a direction which is reverse of the
galaxy main spin cannot be explained with secular evolution processes,
and is a strong evidence for an external origin.

It is thus of great interest to investigate the characteristics of
counter-rotating stars in our merger remnants -- specifically, their
fraction and vertical distribution with radius.  Examples are shown
in Figs.~\ref{counter_nogas} and \ref{counter_gas} for a number of
dissipationless and dissipative retrograde mergers. In these plots
four different regions are selected: (1) at $\mid z\mid\le$1~kpc,
(2) at 1$\le\mid z\mid\le$3~kpc, (3) at 3$\le\mid z\mid\le$5~kpc and
(4) at 5$\le\mid z\mid\le$10~kpc. It is clear that counter-rotating
stars are found at all values of z, from the galaxy midplane 
to 10~kpc above it, and that the probability to find
counter-rotating stars increases with height: the fraction
$N_{ct}/N_{tot}$ of counter-rotating stars to the total number of
stars in the region is $\le$5$\%$ up to $\mid z\mid\le$3~kpc,
typically around 10$\%$ in the 3~kpc$\le\mid z\mid\le$5~kpc and 
it can reach 30$\%$ in the region highest above the disk. This
trend is also due to the fact that the fraction of satellite stars
$N_s$ to the total number of stars increases with z. In other words,
it is the overall increase in the fraction of satellite stars with z
which determines this trend, not simply a decrease in the number of
disk stars as a function of height above the disk. Furthermore the
fraction of counter-rotating stars depends also on the initial
orbital inclination of the satellite: for a satellite on an orbit
inclined by 60 degrees with respect to the primary disk, this
fraction is lower at all z \citep{villalobosH208}.

\begin{figure}
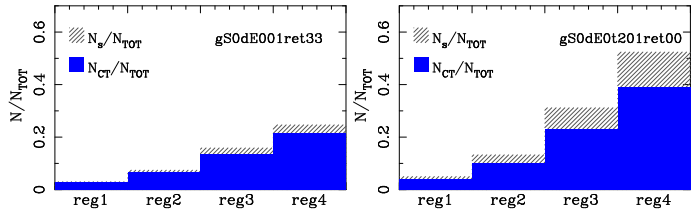

\vspace{0cm}\hspace{-0.2cm}
\centering
\includegraphics[width=2.7cm,angle=270]{16502_49.ps}
\includegraphics[width=2.7cm,angle=270]{16502_50.ps}
\vspace{0.2cm}
\caption{Same as Fig.\ref{counter_nogas}, but for a gS0 galaxy after a
single retrograde merger (left panel) and after two consecutive retrograde mergers (right panel)
with a dE0 satellite.}
\label{multi_counter}
\end{figure}

If counter-rotating stars are a natural outcome of low orbital
inclination in single minor mergers with satellites on retrograde
orbits, would a second retrograde merger then increase the fraction
of counter-rotating stars even further at all heights? In our models
(Fig.~\ref{multi_counter}) the effect of a second retrograde merger
with the same mass ratio is to increase the fraction of
counter-rotating stars by a factor of about two, at all heights. We
can conclude from this that repeated retrograde minor mergers have a
cumulative effect on the number of counter-rotating stars in the
thick disk and inner halo.

\subsection{Kinematics of the remnant disk}

\begin{figure}
\vspace{0.cm}\hspace{0.7cm}
\centering
\includegraphics[width=8.2cm,angle=0]{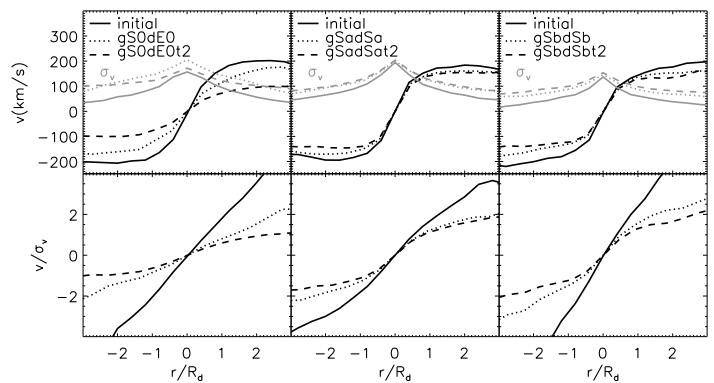}
\vspace{0.7cm} \caption{Line-of-sight velocity, $v$, velocity dispersion, 
$\sigma_v$, and the $v/\sigma_v$ ratio as a function of scaled radius in remnant stellar
disks after both a single and two consecutive dissipative minor mergers.
The same data are also shown for the initial stellar disks, before
merging. The gas mass fraction of the merging galaxies increases from
the left to the right, from 0 for the S0 case, to 0.2 for the Sbs.}
\label{multi-v}
\end{figure}

The AM evolution of the baryonic component during merging processes
discussed in the \S~\ref{angm} is naturally reflected in an evolution
of the line-of-sight velocity $v_{los}$ and the $v_{los}/\sigma_v$ ratio.
Minor mergers always produce a decrease in $v_{los}$ and an increase in the
velocity dispersion $\sigma_v$ of the stellar component at all radii (Fig.~\ref{multi-v}). The
remnant stellar disks usually have a smaller $v_{los}/\sigma_v$ ratio than
the initial values before interaction. Consecutive satellite accretions
have cumulative effects on decreasing the $v_{los}$ and increasing the
$\sigma_v$, and thus deceasing the $v_{los}/\sigma_v$ ratio. Nevertheless,
smaller variations in both $v_{los}$ and $\sigma_v$ can be found in minor
mergers with gas, compared to gas-free merger cases, due to the fact
that dissipative gas can help preserve disk rotation by
forming new rotating disk stars during mergers. The decreases in $v_{los}$
and the $v_{los}/\sigma_v$ ratio indicate that stellar disks become hotter
and rotate more slowly after minor merger events, which is consistent
with the AM loss and the increase of the anisotropy parameter $\beta$
(Figs.~\ref{am} and \ref{multi-am-beta}).

\section{Discussion and Conclusions}\label{discussion}

By means of N-body/SPH simulations, we studied the kinematics of
stars in galaxy remnants of 1:10 mass ratio minor mergers. The
simulated interactions span a range in gas mass fractions, from 0 to
0.2 in the primary and satellite galaxies, and a range of orbital
parameters.

As shown in \citet{quDM210a}, minor mergers result in a
redistribution of orbital into internal angular momentum, which
affects all galaxy components.  In particular, old stars, i.e. those
already in place before the interaction, always lose angular
momentum during the merging process. The decrease of the specific AM
of old stars is accompanied by a redistribution of stellar orbits,
as traced by the anisotropy parameter $\beta$, which become
increasingly radial. In minor mergers with gas in the disk of the
primary galaxy we find a similar trend of old stars losing angular
momentum and as a result their orbits becoming more radially
dominated. However, when a new stellar component forms from gas
present in the primary disk during the merger, its AM content is
significantly different: the orbits tend to be more tangentially
dominated, thus providing a higher rotational support. This
different behavior results in a final stellar disk with two
different stellar populations with significantly different AM
content. In particular, old stars always show a rotational lag with
respect to the young stellar component. If one separates all stars
into thin disk stars (at heights $\mid z\mid\le$1~kpc from the
galaxy midplane) and thick disk stars (at $\mid z\mid >$1~kpc),
three different components can be found, with different dynamical
properties: (1) young stars in the thin disk, which are rotationally
supported and show the highest values of $v_t$, (2) old thin disk
stars lagging with respect to the new stars and (3) old thick disk
stars lagging with respect to both thin disk components. For a minor
1:10 merger, with a satellite accreted on a direct orbit and with an
initial primary disk gas fraction of 0.2, the old stars in the thin
disk have a rotational lag of about 20~km s$^{-1}$, while the old
stars in the thick disk have a velocity about 50~km s$^{-1}$ lower
than the young stellar component, both lag values being compatible
with the estimates for the Milky Way \citep[see ][]{gilmore202}.
Multiple mergers can further reduce the tangential velocity of the
old stellar components, while leaving that of the new stars mostly
unchanged, thus resulting in a further increase in rotational lag
with every successive accretion episode. As the two
populations, old stars and new stars, have different tangential
velocities, in a plot of $v_t$ as function of age we expect to find
a discontinuity at the time when the merger occurs. Of course, the
newly formed stars will evolve in time, and if no other merger takes
place they will be slowly heated by secular effects. However, as we
have shown in Fig.~\ref{totAMisomer}, secular processes are much
less effective in altering stellar kinematics than minor mergers.
Therefore, in our opinion, a discontinuity in the age-$v_t$ plane
(or the age-$\beta$ parameter) between the old and new stellar
populations should still be visible, even if secular processes, and
asymmetric drift in particular, contribute to the slow heating of
the new stellar populations. Unfortunately, the hybrid particle
method we adopted to implement star formation in the simulations
does not allow us to follow properly the heating of the newly formed
stars with time. We do think, however, that this discontinuity in
the age-$v_t$ behavior may be a distinctive feature of minor mergers
compared to secular processes in galaxy disks, which deserves to be
modeled accurately in the near future.

Minor mergers can quantitatively reproduce the increase in the
rotational lag with height above and below the galaxy midplane that
was found for the Milky Way \citep{chibaB200,girard206}, but only if
the satellite is accreted along a direct orbit; for retrograde
orbits no trend of the rotational lag with z is found, up to
distances of 5~kpc from the midplane, above which it even decreases
slightly. Together with the increasing trend of stellar
eccentricities with height \citep{dimatteo210}, this is another
piece of evidence suggesting that if the Milky Way thick disk has
formed through heating of a pre-existing thin disk by minor
merger(s), the orbit of the satellite(s) should have been prograde.

Recently, \citet{bournaud209} proposed a scenario in which stellar
thick disks could have formed in gas-rich galaxies at high redshifts,
through scattering of stars by massive clumps. Comparing the kinematics
of minor merger remnants with those of thick disks formed in
clumpy disks we find that, under certain conditions, both scenarios can
reproduce a rotational lag of the old thin and old thick disk stars that are
compatible with observations for the Milky Way. In particular, disks
heated by massive clumps show a rotational lag which increases with height z,
as found for minor mergers on direct orbits.  Even so, the two
processes leave distinctly different signatures in the vertical surface profiles
of the remnant disks, as shown in \citet{quDM211b}.  Obviously, disks
heated by massive clumps cannot lead to the counter-rotating thick disks
observed in external galaxies \citep{yoachimD208}. For this to occur,
an external accretion of material (gas or stars) is necessary.

Counter-rotating material is found in all our simulations where the
satellite is accreted along a retrograde orbit. The amount of
counter-rotating stars (all associated to stars initially in the
satellite galaxy) increases with z as a result of the increasing
contribution of satellite stars to the whole stellar content, and at
$z=3-5$~kpc the fraction of counter-rotating stars is between 10$\%$
and 15$\%$. We showed that the effect of multiple mergers on the
generation of a counter-rotating disk is cumulative, in the sense
that the fraction of counter-rotating stars doubles after the
accretion of a second, comparatively massive, satellite on a
retrograde orbit.  It is thus possible that some of the galaxies
observed to have a high fraction of counter-rotating material in the
thick disk \citep{yoachimD205} could have accreted a number of
satellites on retrograde orbits. The fraction of counter-rotating
stars we find in our minor merger simulations is quantitatively
consistent with the amount of counter-rotation observed in the thick
disks of nearby galaxies \citep{yoachimD205}.

Finally, we discussed the possibility of discriminating stars
accreted from the satellite galaxy from those formed in the primary
disk on the basis of their AM content, as recently proposed by
\citet{villalobosH209}. We find that in general the AM content is a
better discriminant for stars located at vertical distances from the
galaxy midplane greater than 1~kpc rather than at smaller heights,
as proposed by \citet{villalobosH209}.  Moreover, as the fraction of
satellite stars increases with z in our models, their detection at
high z should also be more likely.  However, we want to point out
that it may not always be possible to use the AM content to
distinguish between stars originally in the primary and those
accreted from the satellite because there are cases where no
difference in their AM content is seen at any radius.

Overall, the qualitative and some quantitative agreement with observations
of the thick disk of the MW and other nearby galaxies suggest that
minor merger remains a viable mechanism for forming the thick disk.
With future observations, it may be possible to search for other signs
of the impact of minor mergers on disk galaxies, such as determining
the relative angular momentum of stars as a function of age in nearby
disk galaxies. Minor merger models, such as those discussed here, make
specific predictions which can be tested observationally (e.g., by GAIA).

\section*{Acknowledgments}
YQ and PDM are supported by a grant from the French Agence Nationale de
la Recherche (ANR). We are grateful to Beno\^it Semelin and Fran\c{c}oise
Combes for developing the code used in this paper and for their permission
to use it.  These simulations will be made available as part of the GalMer
simulation data base (\emph{http://galmer.obspm.fr}). We wish to thank Gary
Mamon for his extensive and useful comments and the anonymous referee for
a careful, in-depth and constructive report which helped improve the paper
considerably.

\bibliographystyle{aa}
\bibliography{bib-paper3}

\end{document}